\documentclass[aps,prb,onecolumn,amsmath,amssymb,superscriptaddress,showpacs]{revtex4}

\usepackage{amssymb}
\usepackage{amsmath}
\usepackage{dsfont}
\usepackage{bbm}
\usepackage{graphicx,epsfig}
\usepackage{wasysym}
\usepackage{float}
\usepackage{color}
\usepackage[section]{placeins}
\usepackage[colorlinks]{hyperref}

\newcommand{\be}{\begin{equation}}
\newcommand{\ee}{\end{equation}}
\newcommand{\up}{\uparrow}
\newcommand{\down}{\downarrow}
\newcommand{\T} {\mathcal{T}}
\newcommand{\I} {\mathcal{I}}

\newcommand{\Ha}{\mathrm{H}}
\newcommand{\ave}{\arrowvert}

\newcommand{\righta}{\rightarrow}
\renewcommand{\dag}{\dagger}
\newcommand{\sigmab}{\bar{\sigma}}

\begin{document}

\title{Superconductor spintronics: modeling spin and charge accumulation in out-of-equilibrium NIS junctions subjected to Zeeman magnetic fields}
\author{D. Chevallier}
\affiliation{Instituut-Lorentz, Universiteit Leiden, P.O. Box 9506, 2300 RA Leiden, The Netherlands}
\author{M. Trif}
\affiliation{Institut de Physique Th\'eorique, CEA/Saclay, Orme des Merisiers, 91190 Gif-sur-Yvette Cedex, France}
\author{C. Dutreix}
\affiliation{Laboratoire de Physique des Solides (CNRS UMR-8502), B\^atiment 510, Universit\'e Paris-Sud/Universit\'e Paris-Saclay, 91405 Orsay Cedex, France}
\affiliation{Univ. Bordeaux, LOMA, UMR 5798, Talence, France and CNRS, LOMA, UMR 5798, Talence, F-33400, France}
\author{M. Guigou}
\affiliation{Institut de Physique Th\'eorique, CEA/Saclay, Orme des Merisiers, 91190 Gif-sur-Yvette Cedex, France}
\affiliation{Laboratoire de Physique des Solides (CNRS UMR-8502), B\^atiment 510, Universit\'e Paris-Sud/Universit\'e Paris-Saclay, 91405 Orsay Cedex, France}
\author{C. H. L. Quay}
\affiliation{Laboratoire de Physique des Solides (CNRS UMR-8502), B\^atiment 510, Universit\'e Paris-Sud/Universit\'e Paris-Saclay, 91405 Orsay Cedex, France}
\author{M. Aprili}
\affiliation{Laboratoire de Physique des Solides (CNRS UMR-8502), B\^atiment 510, Universit\'e Paris-Sud/Universit\'e Paris-Saclay, 91405 Orsay Cedex, France}
\author{C. Bena}
\affiliation{Laboratoire de Physique des Solides (CNRS UMR-8502), B\^atiment 510, Universit\'e Paris-Sud/Universit\'e Paris-Saclay, 91405 Orsay Cedex, France}
\affiliation{Institut de Physique Th\'eorique, Universit\`e Paris Saclay, CEA, CNRS, Orme des Merisiers, F-91191 Gif-sur-Yvette Cedex, France}
\date{\today}

\begin{abstract}
We study the spin and charge accumulation in a superconductor when a normal-superconductor tunnel junction is subjected to a Zeeman magnetic field and taken out of equilibrium by applying either a DC or AC voltage bias. We focus on a configuration which allows one to measure non-locally the spin-accumulation using a second ferromagnetic electrode. Our main result is that, in the presence of an AC bias, the time average of the detected non-local signal is frequency dependent, and the frequency at which the saturation occurs is directly related to the inverse of the spin-relaxation time. 
For a DC bias we address also the effect of the spin accumulation in the normal leads and we investigate the out-of-equilibrium spin susceptibility of the SC, which we show to deviate drastically from its equilibrium value.
\end{abstract}

\maketitle

\section{Introduction}

Superconductors' (SC) potential as spintronics materials is based on the possibility either of manipulating the constituent spins of Cooper pairs in the condensate (to form spin-aligned triplet pairs) or else of spin-polarising quasiparticle excitations.\cite{LinderNatPhys15} Recent experiments have shown a non-equilibirum chargeless spin accumulation in a thin-film mesoscopic superconductor in the presence of a Zeeman magnetic field \cite{quay2013,beckmann2012,beckmann2013} enabling an estimate of the spin relaxation time in mesoscopic superconductors. This is on the order of nanoseconds, significantly longer than the charge-relaxation time \cite{clarke1972,aronov1976,johnson1994,chen2002,chandrasekhar2007,yang2010,clarke1981,quay2013,rokhsar1990,quay2014}. The spin imbalance relaxation length has also been measured up and can be up to 10$\mu$m. \cite{beckmann2012}  These developments have revived interest among theorists in this topic \cite{houzetPRB15,bergeretPRL15,bobkovaPRB15}. 

While taking into account all the microscopic physical effects associated with the spin accumulation (such as the modifications of the spin-dependent density of states, or of the out-of-equilibrium distribution function for each spin, and their spatial dependence,...), may be quite a complex endeavour and can be performed, for instance, by using the Usadel equation for describing the behavior of the Keldysh Green's function as it has been shown in Ref.~\onlinecite{Bergeret2017}, we have chosen to rely  here on a reasonable simplified model which captures the essential physics and enables a good quantitative description of the experimental results. 

Thus we describe the out-of-equilibrium spin imbalance as symmetric shifts in the chemical potentials of the two opposite spins, and we assume a uniform temperature for the whole system. Given the fact that the SC is tunnel-coupled to the ferromagnetic (FM) and the normal-metal probes we neglect the processes of Andreev reflection and we focus solely on the quasiparticle tunnelling processes. We use the Fermi Golden rule to write down the form of the tunnel current  \cite{hershfield1995,tinkham1972,tinkham2004} in the presence of a Zeeman field, which is crucial for transforming the SC into a spin-sensitive material, as it splits its density of states (DOS) for the up and down quasiparticles \cite{tedrow1973,meservey1994}. We compare our results to those obtained from the simpler, semiconductor model for the SC. \cite{tinkham2004}.

Secondly we use the derived form for the spin current and we combine it with time-dependent semiclassical equations of motion for the spin accumulation to calculate the spin accumulation in a SC, in the presence of both an applied DC and AC voltage, for which we calculate also the corresponding time-dependence. We also study a configuration which allows one to measure non-locally such spin-accumulation using a second ferromagnetic electrode.  Moreover we consider second order effects such as the possible spin accumulation in the leads, as well as we calculate the out-of-equilibrium spin susceptibility of the SC, which we show to deviate drastically from its equilibrium value. 

We also consider an applied AC bias and we focus on the time-dependence of the spin accumulation as well as on the possible information that can be obtained from the associated relevant time-scales and frequencies, in particular in what concerns the spin-relaxation time in the SC. Thus our most important result is that the spin-relaxation time can be obtained by examining the dependence of frequency of the measured non-local voltage; we note that such dependence is conditioned by the existence of non-linearities in the detector DOS, here taken to be BCS-like. Our results are qualitatively consistent with a recent experiment \cite{quay2014}. We also focus on the possibility of detecting the difference between a spin-accumulation time and a spin-relaxation time via applying an AC signal made of rectangular pulses.

The paper is organized as follows. In Sec. II we present the theory of spin injection into a SC, and we write down the relation between the injection electric, charge, and spin current as a function of the applied voltage. In Sec. III we present the experimental setup, and we write down the semi-classical equations of motion for the spin. In Sec. IV we solve these equations for a DC bias, and in Sec. V we study sinusoidal and rectangular AC biases. We conclude in section VI.

\section{Tunnelling currents between a SC and a ferromagnet or a normal metal}

In this section we introduce the model used to describe the Ferromagnet$|$Insulator$|$Superconductor (FIS) junction, and we compute the tunnelling currents flowing across it. As described below, in certain limits this formalism also applies to a Normal$|$Insulator$|$Superconductor junction.

\subsection{Theoretical model}

The total Hamiltonian for the junction is : $\Ha=\Ha_F+\Ha_S+\Ha_T$, with $\Ha_{F(S)}$ the Hamiltonian for the ferromagnetic (superconducting) lead, and $\Ha_T$ the tunnelling Hamiltonian between the two. We discuss each term below.

\begin{eqnarray}
\Ha_F&=&\sum_{q,\sigma}\big( \epsilon_{q\sigma}-\mu_F\big) c_{q,\sigma}^{\dag}c_{q,\sigma},
\end{eqnarray}
with $\epsilon_{q\sigma}=\hbar^2  q^2/2 m$ and $\mu_F$ the chemical potential in the FM. 
\begin{eqnarray}
\label{BCSHamiltonian}
\Ha_S&=&\sum_{p,\sigma} \big(\epsilon_p-\mu_s -\sigma \mu_BH \big)c_{p,\sigma}^{\dag}c_{p,\sigma} -\sum_p \Big(\Delta c^{\dag}_{p,\up}c^{\dag}_{-p,\down}+ h. c. \Big) ,
\end{eqnarray}
with $\epsilon_p= \hbar^2p^2/2 m$, $\Delta$ is the superconducting energy gap and $\mu_B H$ is the Zeeman energy.

\begin{eqnarray}
\label{TunnelHamiltonian}
\Ha_T&=&\sum_{p,q,\sigma} \Big(\T_{p,q}^{\sigma}c^{\dag}_{p,\sigma}c_{q,\sigma} + h. c.\Big).
\end{eqnarray}
The first term describes the tunneling of electrons with spin $\sigma$ and momentum $q$, with amplitude $\T_{p,q}^{\sigma}$. We assume that spin but not momentum is conserved in these processes. The ``ferromagnetic'' character of the lead is encoded in the spin-dependent tunnel amplitude. The case of a normal metal can be recovered by making the tunnel amplitude spin independent.

$\Ha_S$ and $\Ha_T$ can be more conveniently expressed in terms of quasiparticle operators. We use Josephson's definition of the Bogoliubov-Valatin transformation\cite{Josephson62,Bardeen61}

\begin{eqnarray}
c^{\dag}_{p,\up}&=&u_p \gamma^{\dag}_{e,p\up}+v_p \gamma_{h,p\down},\\
c^{\dag}_{-p,\down}&=&u_p \gamma^{\dag}_{e,p\down}-v_p \gamma_{h,p\up},
\end{eqnarray}
where the $\gamma^{\dag}_{e(h),p}$ are creation operators of electronlike (holelike) excitations.
Note that, as quasiparticles have probability $u_p^2$ ($v_p^2$) of being an electron (hole), the quasiparticle charge is $q_p=u_p^2-v_p^2$. When an electron tunnels into the superconductor, the corresponding charge carried by the condensate is thus $2v_p^2 =1-q_p$.\cite{hershfield1995}

The Hamiltonian for the superconductor can be written as
\begin{eqnarray}
\Ha_S&=&\mu_s\sum_{p,\sigma}c_{p,\sigma}^{\dag}c_{p,\sigma} + \sum_{p,\sigma}E_{p,\sigma}\Big(\gamma^{\dag}_{e,p\sigma}\gamma_{e,p\sigma}+\gamma^{\dag}_{h,p\sigma}\gamma_{h,p\sigma}\Big),
\end{eqnarray}
where
\begin{eqnarray}
u_p&=&\frac{1}{2}\left(1+\frac{\epsilon_p}{\sqrt{\epsilon_p^2+\Delta^2}}\right),\nonumber\\
v_p&=&\frac{1}{2}\left(1-\frac{\epsilon_p}{\sqrt{\epsilon_p^2+\Delta^2}}\right),\nonumber
\end{eqnarray}
and $E_{p,\sigma}=\sqrt{\epsilon_p^2+\Delta^2}-\sigma \mu_B H$ the excitation energy.\cite{tedrow1973,meservey1994} Introducing $\sigmab=-\sigma$, the tunneling Hamiltonian becomes
\begin{eqnarray}
\Ha_T&=&\sum_{p,q,\sigma} \Big(\T_{p,q}^{\sigma}\big[u_p\gamma^{\dag}_{e,p\sigma}c_{q,\sigma}+v_p\gamma_{h,p\sigmab}c_{q,\sigma}\big] + h. c.\Big).
\end{eqnarray}

\subsection{Tunnel current}

Next we calculate the charge, spin and quasiparticle charge currents flowing through the junction using the Fermi's Golden rule.\cite{maekawa1999,hershfield1995} Table \ref{Processes_Table} depicts the allowed tunneling processes and their corresponding probabilities. For example, the probability for a $\gamma_{e,p\sigma}^{\dagger}c_{q,\sigma}$ process (annihilation of an electron $c_{q,\sigma}$ in the FM and creation of a quasiparticle $\gamma_{e,p\sigma}^{\dagger}$ in the SC) is a product of the tunnel probability given by the tunnel Hamiltonian, $\ave \T_{p,q}^{\sigma} \ave ^2u_{p}^{2}$; the probability of having a filled state in the FM to tunnel from, $f(\epsilon_{q})$; and the probability of having an empty state in the SC to tunnel into, $1-f_{p\sigma}(E_{p,\sigma})$.

\begin{table}[b]
\begin{tabular}{c c c c c c }
~~Process~~ & ~~Probability~~ & ~~Electrons added~~ & ~~Quasiparticle charge~~ & ~~Condensate charge~~ & ~~Spin added~~\\
\hline
$\gamma_{e,p\sigma}^{\dagger}c_{q,\sigma}$ & $\ave \T_{p,q}^{\sigma} \ave ^2 u_{p}^{2}[1-f_{p\sigma}(E_{p,\sigma})]f(\epsilon_{q})$ & $+1$ & $+q_p$ & $1-q_p$ & $\sigma$ \\
$c_{q,\sigma}^{\dagger}\gamma_{e,p\sigma}$ & $\ave \T_{p,q}^{\sigma} \ave ^2 u_{p}^{2}[1-f(\epsilon_{q})]f_{p\sigma}(E_{p,\sigma})$ & $-1$ & $-q_p$ & $-1+q_p$ & $\bar{\sigma}$\\
$\gamma_{h,p\sigmab}c_{q,\sigma}$ & $\ave \T_{p,q}^{\sigma} \ave ^2 v_{p}^{2}f_{p\sigmab}(E_{p,\sigmab})f(\epsilon_{q})$ & $+1$ & $-q_p$ & $1+q_p$ & $\sigma$ \\
$c_{q,\sigma}^{\dagger}\gamma^{\dag}_{h,p\sigmab}$ &$\ave \T_{p,q}^{\sigma} \ave ^2 v_{p}^{2}[1-f_{p\sigmab}(E_{p,\sigmab})][1-f(\epsilon_{q})]$ & $-1$ & $+q_p$ & $-1-q_p$ & $\bar{\sigma}$ \\
\end{tabular}
\caption{Tunneling processes in excitation representation.\cite{hershfield1995,tinkham1972}}
\label{Processes_Table}
\end{table}

The average tunneling current through the junction for a given spin can be written as
\begin{eqnarray}
\I_{\sigma}&=&\frac{2 \pi}{\hbar} \sum_{p,q} \ave \T_{p,q}^{\sigma} \ave ^2 \Big\{u_p^2 \delta(\epsilon_q-E_{p,\sigma}+eV) \big[f(\epsilon_q)-f_{p\sigma}(E_{p,\sigma})\big]-v_p^2\delta(\epsilon_q+E_{p,\sigmab}+eV)\big[1-f(\epsilon_q)-f_{p\sigmab}(E_{p,\sigmab})\big]\Big\},
\end{eqnarray}
where we assumed $\T_{p,q}^{\sigma}=\T_{q,p}^{\sigma}$. To simplify this formula we note that for each state with $p^+>k_F$, energy $E_{p^+,\sigma}$ and $u_{p^+}$, there exists another state $p^-<k_F$ with the same energy $E_{p^-,\sigma}=E_{p^+,\sigma}$ (note however that $\epsilon_{p^+}=-\epsilon_{p^-}$. This implies for the coherence factors that $u_{p^{\pm}}^2=v^2_{p^{\mp}}$. Moreover we can reasonably assume that $\ave \T_{p^+,q}\ave = \ave \T_{p^-,q} \ave$. For the distribution functions in the superconductor, on the other hand, we can write:
\begin{align}
f^{T,C}_{p,\sigma}&=f_{p_+\sigma}(E_{p\sigma})\pm f_{p_-\sigma}(E_{p\sigma})\,,
\end{align}
where $f^{T,\left.(C\right.)}_{p,\sigma}$ is the total (branch imbalanced) quasiparticle distributions for spin $\sigma$. Note that for $f^C_{p,\sigma}=0$ the distribution function on hole and electron branches is the same and thus there is no associated charge imbalance.  By separating the sum over $p$ into two sums over $p^{\pm}$, and by noting that only the coherence factors depend on $p^{\pm}$ (with $u_{p^\pm}=v_{p^\mp}$, $u_{p^\pm}^2+v_{p^\pm}^2=1$) we obtain
\begin{eqnarray}
\I_{\sigma}&=&\frac{2 \pi}{\hbar} \sum_{p^{+},q} \ave \T_{p^+,q}^{\sigma} \ave ^2
\delta(\epsilon_q-E_{p^{+},\sigma}+eV) \big[f(\epsilon_q)-f^T_{p^{+}\sigma}(E_{p^{+},\sigma})\big]\nonumber\\
&-&
\delta(\epsilon_q+E_{p^{+},\sigmab}+eV)\big[1-f(\epsilon_q)-f^T_{p^{+}\sigmab}(E_{p^{+},\sigmab})\big]\Big\}\nonumber\\
&-&\frac{2 \pi}{\hbar} \sum_{p^{+},q} \ave \T_{p^+,q}^{\sigma} \ave ^2q_p^2\left[\delta(\epsilon_q-E_{p^{+},\sigma}+eV)f^C_{p^{+}\sigma}(E_{p^{+},\sigma})+\delta(\epsilon_q+E_{p^{+},\bar{\sigma}}+eV)f^C_{p^{+}\bar{\sigma}}(E_{p^{+},\bar{\sigma}})\right]\,,
\end{eqnarray}
where $q_p=(u_p^2-v_p^2)$ is the quasiparticle charge. Typically, the contribution of the charge imbalance to the currents is negligible, so that we will neglect it in the following. In such a case, the second line in the above expression vanishes (this, however, does not imply that such a current cannot lead to a charge imbalance, a feature discussed later on).  In the following the sign $+$ of the momentum $p$ will be omitted for brevity. We can now convert the momentum summation into an energy integral, using
\begin{eqnarray}
\sum_q \righta \int dq \underbrace{\rho(q)}_{(L/2\pi)^d} \righta \int dE \rho(E).
\end{eqnarray}
For the energy range we consider, it is reasonable to assume that the FM DOS and the tunneling probabilities are roughly independent of energy. Performing the momentum-energy conversion for the $\sum_q$ in the FM, and subsequently the resulting energy integral, we obtain
\begin{eqnarray}
\I_{\sigma}&=&\frac{2 \pi}{\hbar}\sum_p \rho_{F}\ave \T^{\sigma} \ave ^2\left[f(E_{p,\sigma}-eV)-f^T_{p\sigma}(E_{p,\sigma})+f^T_{p\sigmab}(E_{p,\sigmab})-f(E_{p,\sigmab}+eV)\right]\,,
\end{eqnarray}
Here $\rho_{F}$ is the total density of states of the ferromagnetic lead (integrated over the volume of the FM). The conversion of the summation over the momentum $p$ in the SC into an energy integral is more tricky. This is because the two first terms of the above expression correspond to the injection of an electron as an electron-like excitation of energy $E_{p,\sigma}$. The last two terms correspond to the conversion of an electron into a hole-like excitation at energy $-E_{p,\sigmab}$. The SC densities of states are different for the two processes, due to the presence of the Zeeman field: $\rho_S(E_{p,\sigma}) \neq \rho_S(-E_{p,\sigmab})$. Converting the momentum summation over $p$ into an energy integral thus leads to
\begin{eqnarray}
\I_{\sigma}&=&\frac{2 \pi}{\hbar} \rho_{F}\rho_0\ave \T^{\sigma} \ave ^2\int_{-\infty}^{+\infty} dE \Big\{\rho(E_{\sigma})\big[f(E-eV)-f_{\sigma}(E)\big]+\rho(E_{\sigmab})\big[f_{\sigmab}(E)-f(E+eV)\big]\Big\},\nonumber\\
\label{current_per_spin}
\end{eqnarray}
where $E_{\sigma}=E-\sigma \mu_B H$, $\rho_0$  is the DOS of the superconductor at the Fermi energy, and
\begin{eqnarray}\label{dosSC_eq}
\rho(E)&=&\theta(E-\Delta)\frac{E}{\sqrt{E^2-\Delta^2}},
\end{eqnarray}
is the usual normalized BCS density of states, with $\theta(x)$ the Heaviside step function.

The DOS of real superconductors can deviate slightly from the BCS DOS: Magnetic impurities, supercurrents or orbital magnetic fields can round off the BCS singularity (Abrikosov-Gorkov depairing)\cite{abrikosov1961,maki1964,levine1967,tinkham1967,anthore2003}, while states can appear in the gap due e.g. to strong electron-phonon coupling or Andreev reflection at the interface with the (normal) tunnel electrode\cite{blonder1982,dynes1978}.  In numerical calculations, one can also use an experimentally-measured DOS. In the following, we will assume that the departure from equilibrium for all the distribution functions is encoded in Fermi-Dirac distributions with shifted chemical potentials. Under this assumption, the total spin and electric tunneling current can be written
$\I_e=e \sum_{\sigma}\I_{\sigma}$ and $\I_s=(\hbar /2)\sum_{\sigma}\sigma\I_{\sigma}$
where
\begin{eqnarray}
\I_{\sigma}&=&\frac{\pi}{\hbar} \rho_{F}\rho_0\ave \T^{\sigma} \ave ^2\int_{-\infty}^{+\infty} dE \Big\{\rho(E_\sigma )\big[f(E-eV)-f(E-\sigma \mu_s)\big]+\rho(E_{\sigmab})\big[f(E+\sigma\mu_s)-f(E+eV)\big]\Big\}\,,
\label{current}
\end{eqnarray}
where $\mu_s$ refers to the shift of the chemical potential due to the spin imbalance. Explicitly, we obtain for the charge and spin currents:
\begin{eqnarray}
\label{id}
\I_e&=&e\sum_{\sigma}\I_{\sigma},\\
&=&\frac{2 \pi e}{\hbar} \rho_{0}\rho_F\int_{-\infty}^{+\infty} dE\Big\{\big[\ave \T_{\up}\ave ^2\rho(E_\up)+\ave \T_{\down} \ave ^2 \rho(E_\down)\big]f(E-eV)
-\big[\ave \T_{\up}\ave ^2\rho(E_\down)+\ave \T_{\down} \ave ^2 \rho(E_\up)\big]f(E+eV)
\nonumber\\&-&
\big[\ave \T_{\up}\ave ^2-\ave \T_{\down} \ave ^2\big]\big[\rho(E_\up)f(E-\mu_s)-\rho(E_\down)f(E+\mu_s)\big]\Big\},\nonumber
\label{charge_current_FM}
\end{eqnarray}

\begin{eqnarray}
\label{SpinCurrent}
\I_s&=&\frac{\hbar}{2}\sum_{\sigma}\sigma\I_{\sigma}=\pi\rho_0\rho_{F}\int_{-\infty}^{+\infty} dE\Big\{\big[\ave \T_{\up}\ave ^2\rho(E_{\up})-\ave \T_{\down} \ave ^2 \rho(E_{\down})\big]f(E-eV)\nonumber\\
&-&\big[\ave \T_{\up}\ave ^2\rho(E_{\down})-\ave \T_{\down} \ave ^2 \rho(E_{\up})\big]f(E+eV)-\big[\ave \T_{\up}\ave ^2+\ave \T_{\down} \ave ^2\big]\big(\rho(E_{\up})f(E-\mu_s)-\rho(E_{\down})f(E+\mu_s)\big)\Big\}\,,
\label{spin_current_FM}
\end{eqnarray}

Using Table I and the above expressions we can also write down the form for the quasiparticle current:
\begin{eqnarray}
\I_e^{qp}&=&\frac{\pi}{\hbar} \rho_{F}\rho_0\sum_{\sigma}\ave \T_{\sigma} \ave ^2\int_{-\infty}^{+\infty} dE \Big\{q^2(E_\sigma)\rho(E_\sigma)\big[f(E-eV)-f(E-\sigma \mu_s)\big]\nonumber \\
&+&q^2(E_{\sigmab})\rho(E_{\sigmab})\big[f(E+\sigma\mu_s)-f(E+eV)\big]\Big\}\,.
\label{quasiparticle_current}
\end{eqnarray}
where $q(E)=e[u(E)^2-v(E)^2]$ is the quasiparticle charge in the energy domain. 
To understand the meaning of the quasiparticle current, we should note that we take into account only electron tunnelling from and to the SC, and not Andreev reflection processes, thus the charge transfer associated with each tunnelling processes is $e$. However, after entering the SC the electron is being 'decomposed' into a quasiparticle with the same energy (the elementary excitation of the SC), which is a mixture of an electron and a hole, and thus carries a different charge: $q(E)=e[u(E)^2-v(E)^2]$ ($u$ and $v$ the components of the Bogoliubov transformation which correspond to the probability to have an electron or a hole). The rest of the charge of the electron transferred goes into the condensate. See Ref. \onlinecite{Tinkham82} for a comprehensive review. While the charge current has a clear physical significance - the number of the electron injected in the SC per unit of time - the quasiparticle current is a bit more subtle, and is defined only as the number of quasiparticles added to the SC due to the polarization of the junction, per unit of time. So the quasiparticle current is not an actual measurable electrical current, but is rather an artefact current introduced to quantify the dynamics of the quasiparticles in the SC, and thus can be different from the charge current, even in the absence of Andreev reflection. 

In the limit of a normal metal coupled to a superconductor ($\T_{\sigma}\equiv\T$), we obtain:
\begin{eqnarray}
\label{id}
\I_e&=&\frac{2 \pi e}{\hbar} \rho_{0}\rho_F|\T|^2\int_{-\infty}^{+\infty} dE \big[\rho(E_\up)+\rho(E_\down)\big]\big[f(E-eV)-f(E+eV)\big]\,,
\end{eqnarray}
\begin{eqnarray}
\I_s&=&\pi\rho_0\rho_{F}|\T|^2\int_{-\infty}^{+\infty} dE\Big\{\big[\rho(E_{\up})- \rho(E_{\down})\big][f(E-eV)+f(E+eV)]\nonumber\\
&-&2\big[\rho(E_{\up})f(E-\mu_s)-\rho(E_{\down})f(E+\mu_s)\big]\Big\}\,,
\label{spin_current_NM}
\end{eqnarray}

\begin{eqnarray}
\label{current}
\I_e^{qp}&=&\frac{2 \pi e}{\hbar} \rho_{0}\rho_F|\T|^2\int_{-\infty}^{+\infty} dE \big[q^2(E_\up)\rho(E_\up)+q^2(E_\down)\rho(E_\down)\big]\big[f(E-eV)-f(E+eV)\big]\,.
\label{quasi_current_NM}
\end{eqnarray}

\subsection{Relation to the semiconductor model}

We compare the results of the previous section with those obtained from a simplified, `semiconductor' model (SM), in which an electron injected into the superconductor enters as a quasiparticle with the same spin, and consequently accesses solely one spin density of states. (In other words, the superconducting quasiparticles are treated like electrons.) The electron particle current for a given spin in the semiconductor model reads:
\begin{eqnarray}
\I_\sigma^{\rm SM}&=&\frac{\pi}{\hbar} \rho_{F}\rho_0\ave \T_{\sigma} \ave ^2\int_{-\infty}^{+\infty} dE \rho^{SM}(E_\sigma )\big[f(E-eV)-f(E-\sigma \mu_s)\big]\,,
\end{eqnarray}
where
\begin{equation}
\rho^{SM}(E)=\theta(E^2-\Delta^2)\frac{|E|}{\sqrt{E^2-\Delta^2}}\,.
\end{equation}
Note that in the SM the DOS is defined, in the electron representation, for both positive and negative energies, while in our model this is positive-defined in the excitation representation. To compare the results from both models, we rewrite the SM to correspond to the same positive-defined energy as in the excitation model. We start by separating the integral into two:
\begin{align}
\I_\sigma^{\rm SM}&=\frac{\pi}{\hbar} \rho_{F}\rho_0\ave \T_{\sigma} \ave ^2\int_{-\infty}^{\infty} dE\rho(E_\sigma )\big[f(E-eV)-f(E-\sigma \mu_s)\big]+\frac{\pi}{\hbar} \rho_{F}\rho_0\ave \T_{\sigma} \ave ^2\int_{-\infty}^{\infty} dE\rho(-E_\sigma )\big[f(E-eV)-f(E-\sigma \mu_s)\big]\nonumber\\
\end{align}
and obtain
\begin{align}
\I_\sigma^{\rm SM}&=\frac{\pi}{\hbar} \rho_{F}\rho_0\ave \T_{\sigma} \ave ^2\int_{-\infty}^{\infty} dE\left\{\rho(E_\sigma )\big[f(E-eV)-f(E-\sigma \mu_s)\big]+\rho(E_{\bar{\sigma}})\big[f(E+\sigma \mu_s)-f(E+eV)\big]\right\}\,,
\end{align}
which is precisely the expression for the current obtained from the excitation model. We note that in deriving this expression, we made the change of variables $E\rightarrow-E$ which for the quasiparticles energies means $E_{\sigma}\rightarrow-E_{\bar{\sigma}}$, since $E_\sigma=E-\sigma E_Z$ (with $E_Z=\mu_B H$). However, in the expression for the current only $|E_{\sigma}|$ matters, so that $|E_{\sigma}|\rightarrow|E_{\bar{\sigma}}|$ and this is how the spin density of states $\rho_{E_{\bar{\sigma}}}$ appears in the expression for the current $\I_{\sigma}$. Such a comparison holds only in the case of zero charge imbalance.

In sum, the two models give identical results for electrical and spin currents; however, the former cannot properly account for imbalances in quasiparticle number nor for quasiparticle charge; a careful calculation of these quantities requires the excitation model described in the previous section.

\section{Charge and spin accumulations}

\begin{figure}[ht]
\centering
		\includegraphics[width=0.7\linewidth]{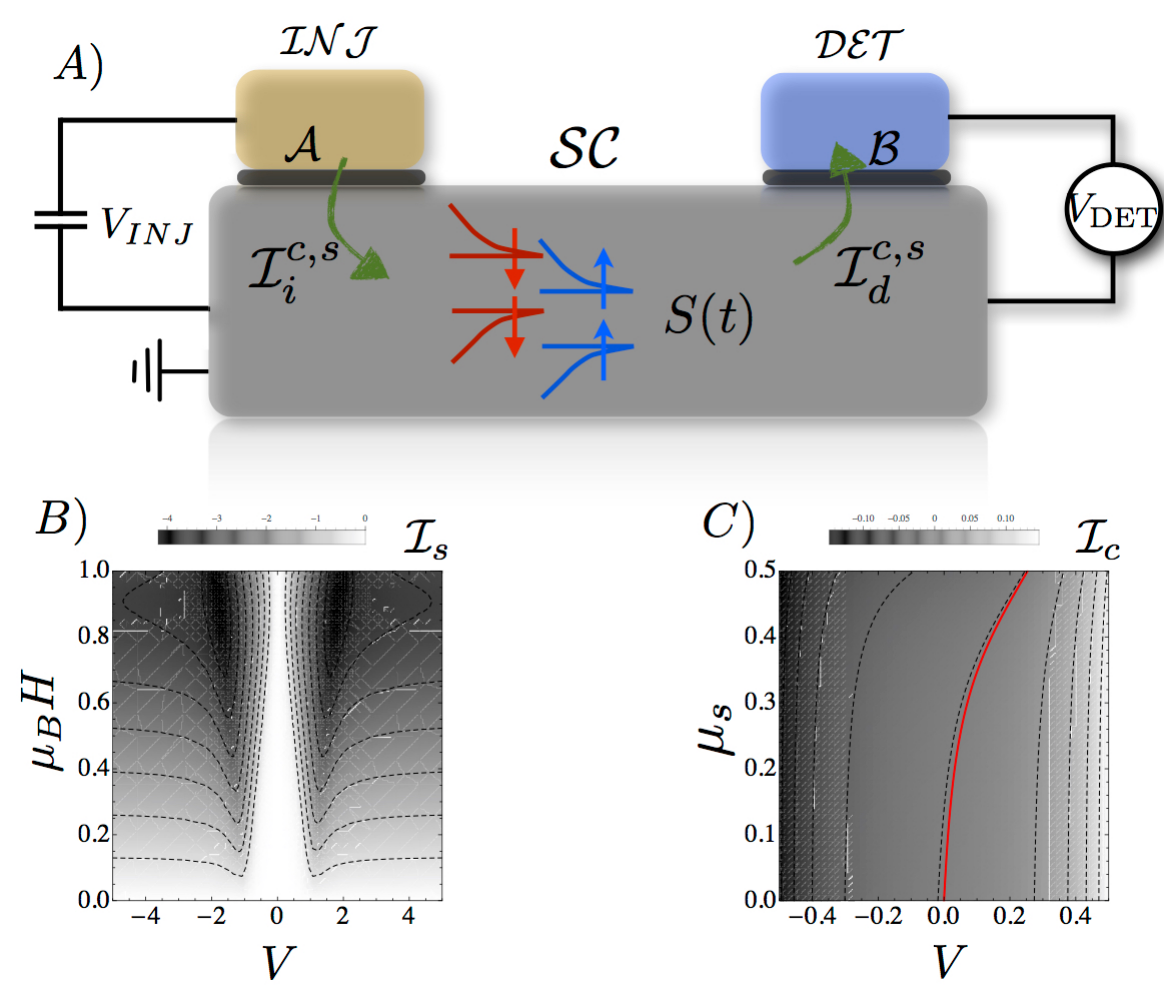}
		\caption{(Color online) A) The experimental setup: The left junction between a normal metal and the Zeeman-split superconductor is voltage biased, which gives rise to charge and spin currents. The detection, on the other hand, is performed with a ferromagnet.  B) The spin  current (in arbitrary units) at the normal (left) injector as a function of the applied voltage $V$ and Zeeman splitting $E_Z=\mu_BH$. C) The charge current flowing into the ferromagnet detector (in arbitrary units), as a function of voltage and spin accumulation $\mu_S$, along with the voltage bias  (in red) corresponding to  the condition of zero current. All quantities are expressed in terms of the SC gap energy $\Delta$}
	\label{fig:Fig0}
\end{figure}

In this section we investigate the spin and charge currents, as well as the resulting spin accumulation in the superconductor and the detector signal --- both as a function of time (in the case of a time-varying injector voltage) and in the steady state where the injector voltage is time-independent.

We consider the non-local setup in Fig. \ref{fig:Fig0}~A. A voltage bias $V$ is applied between a ferromagnetic (or normal) lead and the SC at point A (injection junction). This voltage drop is accompanied by charge and spin currents that can lead to spin accumulation $S(t)$ in the superconductor. The spin current is finite even for a normal injector, as shown in Fig.~\ref{fig:Fig0} B, but vanishes at zero Zeeman splitting. $S(t)$ cannot be detected locally, as the charge current at the injection junction is insensitive to this quantity (see Fig.~\ref{fig:Fig0}~C); however it can be measured nonlocally, as a voltage $V_{DET}$ between a ferromagnetic lead and the superconductor at point B as shown in Fig.~\ref{fig:Fig0} A. Note that both injection and detection junctions obey the same equations, but with different parameters and boundary conditions.

\subsection{Semiclassical equations of motion for the spin imbalance}

We assume that the (total) time dependent spin accumulation $S(t)$ in the superconductor satisfies a simple equation of motion
\begin{eqnarray}\label{motion_eq}
\frac{d S(t)}{dt}&=&\I^i_s(t)-\frac{S(t)}{\tau_s},
\end{eqnarray}
where $\tau_s$ is the spin relaxation time in the SC.
As described in the previous section, $\I^i_s$, the spin current in the injection junction, is a function of the applied voltage $V$ between the SC and the injection lead. Unless otherwise stated, the injection lead is a normal metal (i.e. $\T^i_{\down}=\T^i_{\up}=\T^i$), so that the spin current is given by Eq.~\eqref{spin_current_NM}. Moreover, we assume infinite diffusion length in the superconductor, which is a good approximation for samples of the order of micrometer, as found experimentally. 

The out-of-equilibrium spin accumulation  in the superconductor can be written as
\begin{align}\label{accu_eq}
S(t)&=(\hbar/2)\sum_{k}[f_{k\uparrow}(E_{k,\uparrow})-f_{k\downarrow}(E_{k,\down})]-S_{eq}=(\hbar/2) \rho_0 \int dE\left[\rho(E_{\down})f_{\down}(E)-\rho(E_{\up})f_{\up}(E)\right]-S_{eq}\nonumber\\
&=(\hbar/2) \rho_0 \int dE \left[\rho(E_\down)f(E+\mu_s)-\rho(E_\up)f(E-\mu_s)\right]-S_{eq}\,,
\end{align}
where
\begin{equation}
S_{eq}=(\hbar/2) \rho_0 \int dE \left[\rho(E_\down)-\rho(E_\up)\right]f_0(E)\,,
\end{equation}
is the  equilibrium (thermodynamic) magnetization (which does not contribute to the electronic signal). Note that this equation can be written in such a way only for temperature very low where the Fermi function $f_0(E)$ can be modeled by a step function. We can write the solution for the spin accumulation $S(t)$ as follows:
\begin{eqnarray}\label{imbalance_eq}
S(t)&=&e^{-t/\tau_s} \int_0^{t} dt'\I^i_s(t') e^{t'/\tau_s}.
\end{eqnarray}
The equations (\ref{spin_current_NM}), (\ref{accu_eq}), and (\ref{imbalance_eq})  form a self-consistent system of integral equations which can be solved numerically to determine $\mu_s$ as a function of $V$ for all times $t$. (Note that $\I^i_s$ is also a function of $\mu_s$.) For very small values of $\mu_s$ we can neglect the dependence of $\I^i_s$ on $\mu_s$ and calculate $S(t)$ and $\mu_s$ directly from Eq.~(\ref{imbalance_eq}), as was done in Ref.~\onlinecite{quay2013}.

In the case of a time-independent injection voltage, a dynamic equilibrium appears between the injected spin current and the the spin relaxation in the superconductor, such that $dS(t)/dt=0$. Imposing this condition yields $\I^i_s=S/\tau_s$, with both $S$ and $\I^i_s$ time-independent. Equations (\ref{spin_current_NM}) and (\ref{accu_eq}) then form a self-consistent system of equations which can be solved numerically to determine $\mu_s$ as a function of $V$.

The detector voltage $V_{DET}$ as a function of $\mu_s$ (or $V$) is determined by imposing the condition of zero total electrical current $\I^d_e=0$ at the detector junction. ($\I^d_e$ is given in Eq~\eqref{charge_current_FM}, with the tunneling parameters corresponding to the detector ferromagnet.) Here, the ferromagnetic character of the detector ($\T_{\up}^d \ne \T_{\down}^d$) is crucial; were for non-ferromagnetic detectors, $V_{DET}=0$ for all accumulated $\mu_s$. Conversely, the measured $V_{DET}$ value depends strongly on the polarization of the detector $P^d=(|\T^d_\uparrow|^2-|\T^d_\downarrow|^2)/(|\T^d_\uparrow|^2+|\T^d_\downarrow|^2)$, with a typical experimental $P^d\approx10\%$ for cobalt.

\subsection{Charge imbalance in a Zeeman split superconductor}

To obtain the charge imbalance, we first calculate the quasiparticle charge current using Eq.~\eqref{quasiparticle_current}:
\begin{eqnarray}
\I_e^{qp} &=& \frac{2 \pi e}{\hbar} \rho_{F}\rho_N \ave \T\ave ^2 \int_{-\infty}^{+\infty} dE_{p} \big[q^2(E_{\up})\rho(E_{\up})+q^2(E_{\down})\rho(E_{\down})\big]\big[f(E-eV)-f(E+eV)\big]\,,
\end{eqnarray}
and therefore the charge imbalance (induced by  $\I_e^{qp}$) is independent of the spin imbalance 
We assume that the total quasiparticle charge accumulation $Q(t)$ in the superconductor  obeys a similar equation of motion as for the the spin accumulation:
\begin{equation}
\frac{d Q(t)}{dt}=\I_{e}^{qp}(t)-\frac{Q(t)}{\tau_{Q}},
\end{equation}
where $\tau_Q$ is the charge relaxation time in the SC. Note that the superconductor as a whole stays neutral as the condensate absorbs the charge difference.

The quasiparticle charge accumulation in the superconductor can be written
\begin{align}
Q(t)&=e\sum_{k,\sigma}q_k[f_{k\sigma}(E_{k\sigma})-f_0(E_{k\sigma})]\equiv e\sum_{k+_,\sigma}q_{k_+}f^C_{k_{+}\sigma}(E_{k_{+}\sigma})\nonumber\\
&=\frac{2\pi e}{\hbar}\sum_{\sigma}\int_{-\infty}^\infty q(E_\sigma)\rho(E_\sigma)f^C_{\sigma}(E)\,,
\end{align}
where we recall that $f^C_{\sigma}(E)$ is the distribution function imbalance for spin orientation $\sigma$ and $f_0(E)$ the Fermi-Dirac distribution function. The latter quantity is symmetric with respect to the branch index; thus, its effect vanishes (since $q_k$ is odd). 

\subsection{Spin accumulation in the leads}

In the previous sections, we focused on the superconductor, assuming that the leads were 'inert'. Here we extend our theory to include spin accumulation in the leads (normal and/or ferromagnetic) and its influence on the system as a whole.

The non-local voltage detection of the spin accumulation in the superconductor was assumed to be performed well within the relaxation length $\lambda_{SC}$, and thus we neglected the spatial dependence of the spin accumulation. However, in the normal metals the spin relaxation time $\tau_N$ (and thus the diffusion length $\lambda_{N}$) is much shorter, and we are obliged to use the full diffusion equation in order to properly describe the resulting spin accumulation. For generality, we write down the diffusion equation corresponding to a ferromagnet, and take the limit of the normal metal when necessary. This reads:
\begin{equation}
\frac{\partial n_{\sigma}^\alpha}{\partial t}=D_\sigma^\alpha\nabla^2n_\sigma^\alpha-\left(\frac{n_\sigma^\alpha}{\tau_\sigma^\alpha}-\frac{n_{\bar{\sigma}}^\alpha}{\tau_{\bar{\sigma}}^\alpha}\right)\,,
\end{equation}
where $n_{\sigma}^\alpha\equiv n_{\sigma}^\alpha(x,t) $, $D_{\sigma}^\alpha$, and $\tau_{\sigma}^\alpha$ are the out-of-equilibrium electronic population, diffusion constant, and relaxation time, respectively,  for spin orientation $\sigma$ in lead $\alpha=INJ,DET$. From equilibrium analysis, one can infer that $\rho_\downarrow^\alpha/\tau_\downarrow^\alpha=\rho_\uparrow^\alpha/\tau_\uparrow^\alpha$.  These equations need to be supplemented by boundary conditions at the interface with the superconductor, which here implies that the diffusion current equals the spin current over the interface:
\begin{equation}
D_\sigma^\alpha\nabla n_\sigma^\alpha=-\mathcal{I}_\sigma^\alpha\,,
\end{equation}
where $\mathcal{I}_\sigma^\alpha$ is the current in lead $\alpha$ and for spin $\sigma$, previously calculated (Eq.~\eqref{current_per_spin}). Solving the one-dimensional diffusion equations  for the spin accumulation in the stationary regime $\partial n_{\sigma}^\alpha/\partial t=0$, we obtain:
\begin{align}
n_\sigma^\alpha(x)&=a_\sigma^\alpha e^{-x/\lambda_{D}^\alpha}+b_\sigma^\alpha e^{x/\lambda_D^\alpha}\,,
\end{align}
where $\lambda_D^\alpha=\lambda_\uparrow^\alpha\lambda_\downarrow^\alpha/\sqrt{(\lambda_\uparrow^\alpha)^2+(\lambda_\downarrow^\alpha)^2}$, with $\lambda_{\sigma}^\alpha=\sqrt{D_{\sigma}^\alpha\tau_\sigma^\alpha}$ being an effective diffusion length, and $a_\sigma^\alpha$ and $b_\sigma^\alpha$ are  coefficients are to be found from the boundary conditions (BCs). Assuming that at infinity $n_\sigma^\alpha(-\infty)=0$, we get that $a_\sigma^\alpha=0$. Moreover, we obtain that $D_{\uparrow}^\alpha n_\uparrow^\alpha+D_{\downarrow}^\alpha n_\downarrow^\alpha=0$. From the conservation of currents at the interface between the  lead $\alpha$ and the superconductor (i. e. $x=0$) we get that  $b_\sigma^\alpha=-(\lambda_D^\alpha/D_\sigma^\alpha)\mathcal{T}_\sigma^\alpha$.
The relation between the spin density $n_\sigma^\alpha(x)$ and the resulting chemical potential $\mu_\sigma^\alpha(x)$ is given by $\mu_\sigma^\alpha(x)=n_\sigma^\alpha(x)/\rho_\sigma^\alpha$, so that   we can relate the spin chemical potentials in the lead $\alpha$ at $x=0$ to the corresponding spin currents as follows:
\begin{equation}
\mu_\sigma^\alpha(0)=-\frac{\lambda_D^\alpha}{\rho_\sigma^\alpha D_\sigma^\alpha}\mathcal{I}_\sigma^\alpha=-\frac{\tau_\sigma^\alpha\lambda_\sigma^\alpha}{\rho_\sigma^\alpha(\lambda_D^\alpha)^2}\mathcal{I}_\sigma^\alpha\,.
\end{equation}
The currents $\mathcal{I}_\sigma^\alpha$ themselves depend on the chemical potentials $\mu_\sigma^\alpha(0)$ (via the Fermi's Golden rule), and thus we need to find them self-consistently. The chemical potential can be written as:
\begin{equation}
\mu_\sigma^\alpha=\mu_c^\alpha+\sigma\mu_s^\alpha\,,
\end{equation}
where $\mu_c^\alpha$ and $\mu_s^\alpha$ stand for the charge and spin chemical potentials, respectively. For the currents on the other hand we can write:
\begin{align}
\mathcal{I}_\sigma^\alpha&=\frac{\pi}{\hbar} \rho_{F}\rho_0\ave \T^{\sigma} \ave ^2\int_{-\infty}^{+\infty} dE \Big\{\rho(E_\sigma )\big[f(E-eV_\alpha-\sigma\mu_s^\alpha)-f(E-\sigma \mu_s)\big]\nonumber\\
+&\rho(E_{\sigmab})\big[f(E+\sigma\mu_s)-f(E+eV_\alpha+\sigma\mu_s^\alpha)\big]\Big\}\,,
\end{align}
where $V_\alpha$ is the sum of the applied and the induced voltages over contact $\alpha$. There are thus two more parameters $\mu_{s}^{INJ,DET}$ that need to be found, along with the spin accumulation $\mu_s$ in the superconductor and  $V_{DET}$ the voltage drop on the second contact when the condition of zero current current $I_c^{DET}=0$ is imposed. There are a total of five equations, three of which were already put forward in the previous section [Eqs.~(\ref{spin_current_NM}), (\ref{accu_eq}), and (\ref{imbalance_eq})], while the  two new ones  read:
\begin{align}
\mu_s^\alpha(0)&=-\sum_{\sigma}\sigma\frac{\tau_\sigma^\alpha\lambda_\sigma^\alpha}{\rho_\sigma^\alpha(\lambda_D^\alpha)^2}\mathcal{I}_\sigma^\alpha\,,
\end{align}
where $\alpha=INJ,DET$. Note that assuming $\tau_{\sigma}^\alpha=0$ reduces to the case studied in the previous section, without any spin accumulation in the leads. Moreover, even when the two leads are perfect spin sinks, and thus $\mu_{s}^{INJ,DET}=0$, there is a contribution from the detector to $\mu_s$; however, this can be neglected when $V_{DET}\ll V$, as it is usually the case.

It is worth mentioning that the spin accumulation in the leads can affect the charge imbalance in the superconductor. To see that, let us write explicitly the quasiparticle charge  current in the presence of  spin accumulation in the leads:
\begin{eqnarray}
\label{id}
\I_e^{qp}=\frac{2 \pi e}{\hbar} \rho_{0}\rho_N\ave \T\ave ^2\sum_\sigma\int_{-\infty}^{+\infty} dEq^2(E_\sigma)\rho(E_\sigma)[f(E-eV-\sigma\mu_{INJ})-f(E+eV-\sigma\mu_{INJ})]\,,
\label{charge_current_FM}
\end{eqnarray}
which, as stated, leads to a dependence of the charge imbalance on the spin accumulation in the normal lead. However,  the detailed study of such dependence is beyond the scope of this paper.

In what follows, unless otherwise stated, the injector is a normal metal, and the detector is a ferromagnet with a variable polarization (which can be zero).

\section{Results for an applied DC voltage}

\begin{figure}[ht]
	\centering
		\includegraphics[width=0.47\linewidth]{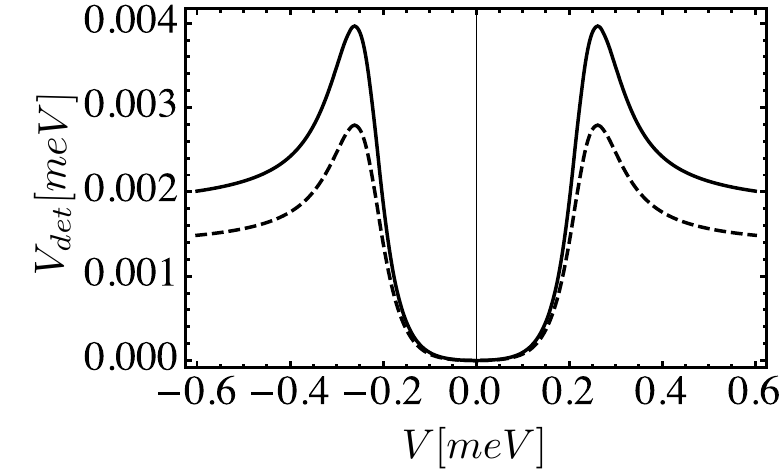}\hspace{0.5cm}
		\includegraphics[width=0.45\linewidth]{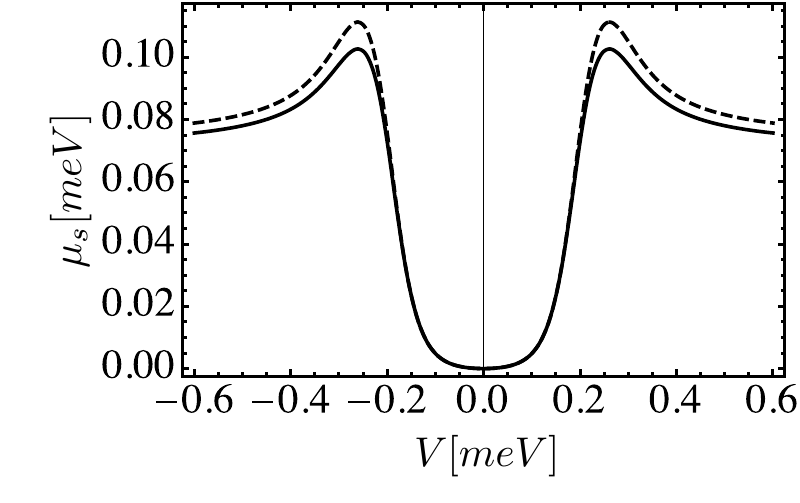}
		\caption{Left: The detector voltage  $V_{DET}$ (in meV) as a function of the applied chemical potential $V$ (also in $meV$) calculated in a self-consistent manner (full line) and non-self consistent manner (dashed line) for parameter values of $\Delta=0.22$ meV, $P^d=2\%$,and $\tau_s=0.1$ in the normalized units described in the text. The DOS is considered to be of BCS Dynes type with a $\delta_i=7$ $\mu$V for the injection DOS, $\delta_d=1$ $\mu$V for the detector DOS (experimentally the DOS inside the SC can be inhomogenous and differ between the injection and detection points, as noted in Ref.~\onlinecite{quay2014}). Right: The calculated accumulated spin-chemical potential $\mu_s$ as a function of the applied voltage using the self-consistent formalism (full line), and the non-self-consistent formalism [multiplied by a factor of $1/(1-2 \tau_s)$] (dashed line). The temperature is taken to $T=270$ mK. The magnetic field is $H=0.2$T.}
	\label{fig:Fig1}
\end{figure}

In Fig.~\ref{fig:Fig1} we present a typical dependence of $V_{DET}$ on $V$. Similar to Ref.~\onlinecite{quay2013} this follows qualitatively the form of $\I_s(V)$ which exhibits the same main features as the BCS DOS \cite{quay2013} (e.g. two coherence peaks, a null value at small $V$'s and a saturation at large $V$'s). Note that we do not take into account the renormalization of the superconducting gap. This would be interesting to explore, but is beyond the scope of this work.
Below we take into account separately the effects of two important factors, self-consistency and non-linearity of the detector junction. This will allow us to understand the difference between our approach and previous approximations. Self-consistency take into account the back-action of the accumulated $\mu_s$ on the injection current; this is negligible when $\mu_s$ is small with respect to the applied voltage and can be neglected in Eq.~(\ref{spin_current_NM}). The detector non-linearity comes into play when $H$ and $\mu_s$ are large, and the above current formulae cannot be Taylor expanded in these parameters. \\

\subsection{Self-consistency}

In the left plot in Fig.~\ref{fig:Fig1} we present the dependence of $V_{DET}$ on $V$ obtained both self-consistently (full line) and non-self-consistently (dashed line). It would appear that the main difference is quantitative, i.e. the self-consistency introduces an overall correcting factor which does not depend strongly on $V$.

To check this, we write down the self-consistent and non-self-consistent solutions of the equations of motion in the linear limit (small $\mu_s$).  In this regime we can perform a Taylor expansion of Eq.~(\ref{accu_eq}) in $\mu_S$, $S= \hbar \rho_0 \rho(\mu_BH) \mu^{nsc}_s$, which together with the condition  $\I^i_s=S/\tau_s$ yields
\be
\mu_s^{nsc}=\tau_s[\hbar  \rho_0 \rho(\mu_BH)]^{-1} \I^i_s(V).
\ee
We can also solve the equations of motion self-consistently by making a Taylor expansion of Eq.~(\ref{spin_current_NM}) in $\mu_s$
\be
\I^i_s=\I^i_s(V)+(2 g_{ns}\hbar/e^2) \rho(\mu_B H) \mu^{sc}_s =S/\tau_s,
\ee
with $g_{ns}=(2 \pi e^2 /\hbar) \rho_F |\T^i|^2  \rho_0$ is the normalized conductance of the injection junction. Noting that $S=2 \hbar \rho_0 \rho(\mu_BH)  \mu^{sc}_s$, we find
\be
\mu_s^{sc}= \frac{\tau_s [\hbar \rho_0 \rho(\mu_BH) ]^{-1}\I^i_s(V) }{1- 2\tau_s g_{ns}/e^2 \rho_0}
\ee
and $\mu^{sc}_s/\mu^{nsc}_s=1/(1- 2\tau_s g_{ns}/e^2 \rho_0)$.
In our numerical calculations we will set $g_{ns}/e^2 \rho_0=1$, so that $\tau_s$ is measured in units of $e^2 \rho_0/g_{ns}$.
Indeed, it seems that in the linear limit, the non-self-consistent and self-consistent approaches differ by a simple numerical factor, which converges to $1$ when $\tau_s \ll 1$. This observations has been checked numerically in the right plot in  Fig.~\ref{fig:Fig1} where we have plotted the accumulated $\mu_s$ calculated using the non-self-consistent and the self-consistent approach (with a correcting factor of $1/(1-2 \tau_s)$ taken into account). Indeed we see that the two give the same result in the linear (small $V$) regime.

It appears thus that the effect of solving the equations of motion in a self-consistent or non-self-consistent manner is mainly quantitative (an overall numerical factor), which is however very important if we are interested in extracting the value of the spin relaxation time from a fit of the experimental data for an applied DC voltage. \\

\subsection{Detector non-linearity}

\begin{figure}[ht]
	\centering
		\includegraphics[width=7cm]{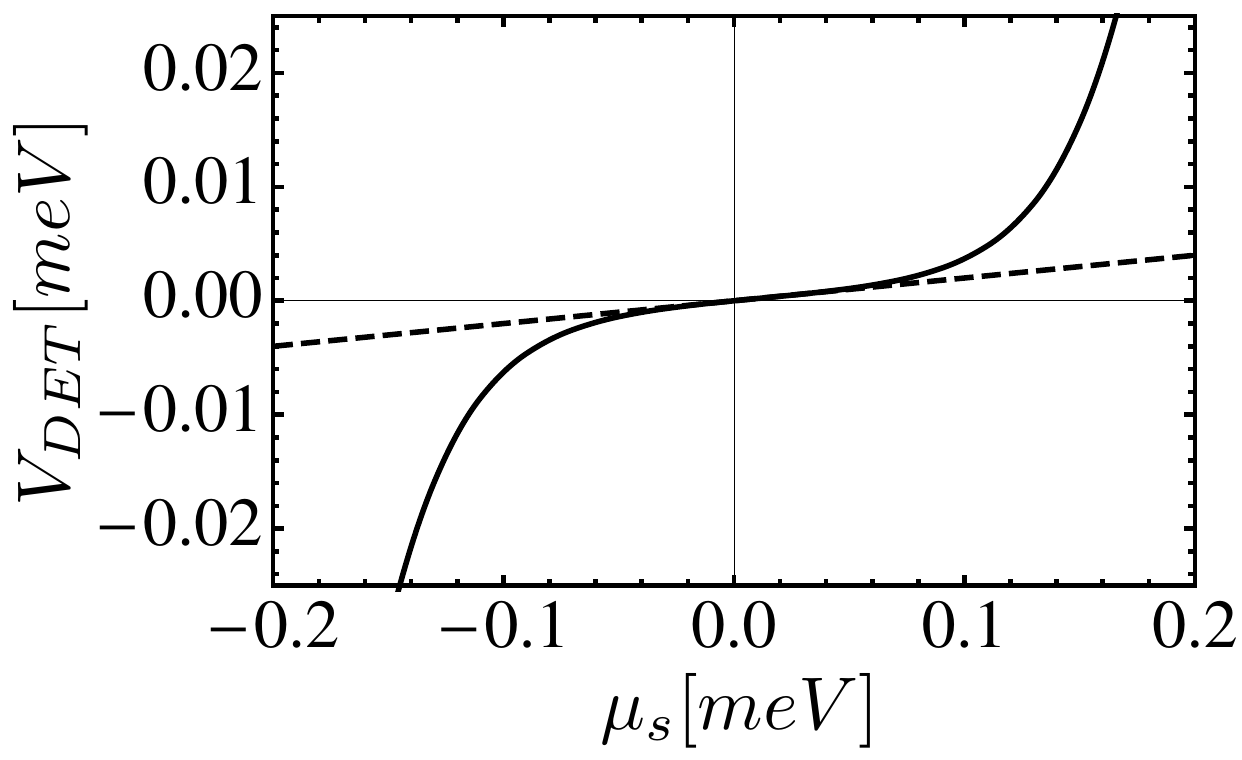}
		\caption{The measured voltage $V_{DET}$  (in $meV$) as a function of $\mu_s$ (full line). The dashed line corresponds to $V_{DET}=P^d \mu_s$. The parameters are the same in Fig.~\ref{fig:Fig1}.}
	\label{fig:Fig3}
\end{figure}

Both the chemical potential describing the spin accumulation $\mu_s$ and the detector voltage $V_{DET}$ depend qualitatively on the injector voltage $V$ in a similar way; any differences come from the non-linearity of the detection junction. To understand this, we have plotted in Fig. \ref{fig:Fig3} the measured $V_{DET}$ as a function of the accumulated $\mu_s$.
Note the pronounced non-linearity of the detection junction, thus for $V\leq \Delta/2$, corresponding to a small $\mu_s$, we have a linear dependence of $V_{DET}$ on $\mu_s$, $V_{DET}\propto P^d \mu_s$ as expected, while for $V > \Delta/2$  the linearity does not hold (as a reminder: $\Delta=0.22$ meV and $P^d=2\%$). We should also note that, as we will show in the next section, the non-linearities in the system are a crucial ingredient in observing a frequency dependence of the measured non-local signal, i.e $V_{DET}$ depends on the frequency of the applied AC voltage only because of the non-linear form of the SC DOS.\\

\subsection{The effect of spin accumulation in the leads}

\begin{figure}[t]
\begin{center}
\includegraphics[width=0.9\linewidth]{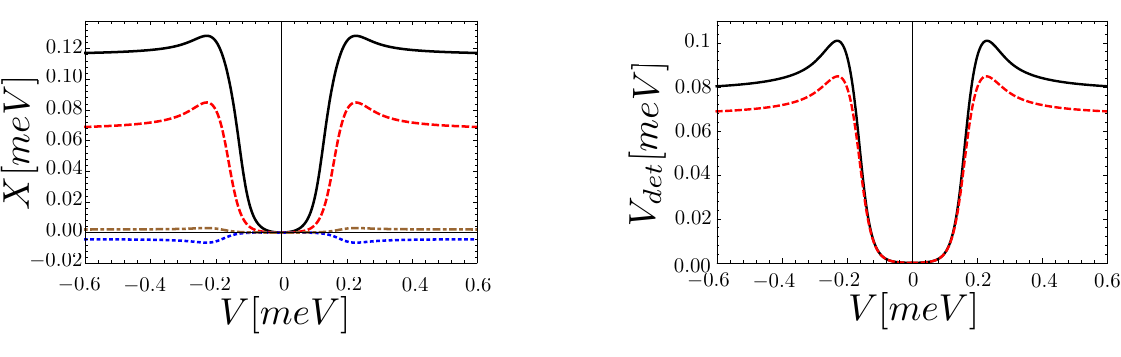}
\caption{The spin accumulations and non-local voltage as a function of the applied voltage at the injection $V$. Left:  $X=\mu_{INJ},\mu_{DET},\mu_s$, and $V_{DET}$ as a function of the voltage $V$. The black (full), red (dashed), blue (dotted), and brown (dot-dashed) curves correspond to $\mu_s$, $V_{DET}$, $\mu_{INJ}$, and $\mu_{DET}$, respectively. Right: the induced non-local voltage $V_{DET}$ when neglecting the spin accumulation in the leads (black-full curve) and when taking into account such accumulations (red-dashed curve).    All energies are expressed in terms of the superconducting gap $\Delta$, and we assumed $H=0.38$T, $T=220mK$, $P=20\%$ (polarization of the right lead).}
\label{FullPotentials}
\end{center}
\end{figure}

Physically, spin accumulation in the leads (parametrized by $\mu_{INJ}$ and $\mu_{DET}$), which has an opposite sign to that in the superconductor ($\mu_s$) will tend reduce the latter; one can think of this as spin accumulation in the superconductor `leaking out' to the leads. Including $\mu_{INJ}$ and $\mu_{DET}$ increases the complexity of the problem drastically --- we now have four equations with four unknowns. We solve the equations numerically, using the same values for the spin relaxation in the superconductor as before, and assume $\tau_\sigma^\alpha\equiv\tau_N=0.1$. In  Fig.~\ref{FullPotentials} we plot the resulting spin accumulations and non-local voltage as a function of the applied voltage $V$ (left plot) and the comparison between the induced voltage $V_{DET}$ when neglecting the accumulation in the leads, and when such accumulations are taken into account.

We see that the voltage becomes slightly reduced as compared to the approximate result, as some of the spin accumulation leaks into the leads. However, for short spin relaxation times in the leads  $\tau_N\ll\tau_S$ in the leads, we can safely neglect such effects, and we proceed with this approximation in the following sections.

We already proved that for accessing the spin accumulation in the superconductor one needs a non-local type of measurement, as the local conductance measurement is independent of such accumulation. However, that is not the case of the spin accumulation in the leads, which is revealed in such local measurements. Specifically,  we mention that the charge current in such a case is given by:
\begin{eqnarray}
\label{id}
\I_e=e\sum_{\sigma}\I_{\sigma}=\frac{2 \pi e}{\hbar} \rho_{0}\rho_N\ave \T\ave ^2\sum_\sigma\int_{-\infty}^{+\infty} dE\rho(E_\sigma)[f(E-eV-\sigma\mu_{INJ})-f(E+eV-\sigma\mu_{INJ})]\,,
\label{charge_current_FM}
\end{eqnarray}
where it appears clearly  that the spin accumulation $\mu_s$ in the superconductor dropped out of the formula. However, the spin accumulation in the normal lead is still present in the expression for the charge current, and leads to a change in the differential conductance
\begin{equation}
G_{NS}(V,\mu_{INJ})=\frac{d\mathcal{I}_{e}}{dV}\,.
\end{equation}
\begin{figure}[t]
\begin{center}
\includegraphics[width=0.9\linewidth]{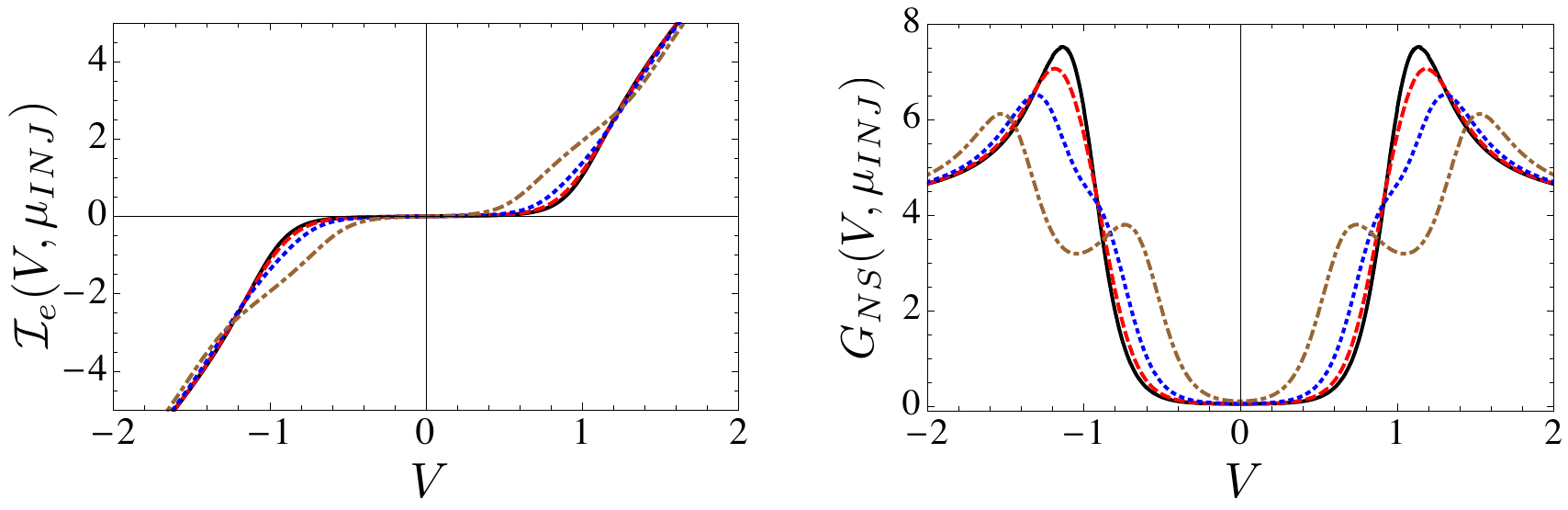}
\caption{The charge current (left) and the differential conductance (right) in arbitrary units as a function of the applied voltage for various values of the spin chemical potential $\mu_{INJ}$ in the normal lead. The black, red, blue, and brown curves correspond to $\mu_{INJ}=0$, $0.1$, $0.2$, and $0.4$, respectively. We assume $E_Z=0$ for all the plots, and all energies are expressed in terms of the superconducting gap energy $\Delta$. }
\label{current_voltage}
\end{center}
\end{figure}

We mention also that if the superconductor is in the normal state, the spin accumulation also drops out of the final expression after the integration. In Fig.~\ref{current_voltage}  we plot the charge current (left) and the differential conductance (right), respectively,  as a function of the applied bias for different values of the spin accumulation $\mu_{INJ}$.  We assume, for simplicity, $E_Z=0$ for all plots. However, the effect of the magnetic field  is trivial: it can be taken into account by only shifting the spin chemical potentials $\mu_S\rightarrow\mu_S-E_Z$, with $S=INJ,s$ (that is only the case when the imbalance is taken into account as just shifting the equilibrium distribution functions. For more complicated distribution functions, such a identification should not hold). There is a clear dependence of both the current and the differential conductance on the spin accumulation, especially when the voltage bias is comparable to the superconducting gap $\Delta$ (note that all energies are expressed in terms of this scale).

\subsection{Out-of-equilibrium spin susceptibility}

\begin{figure}[ht]
	\centering
		\includegraphics[width=0.9\linewidth]{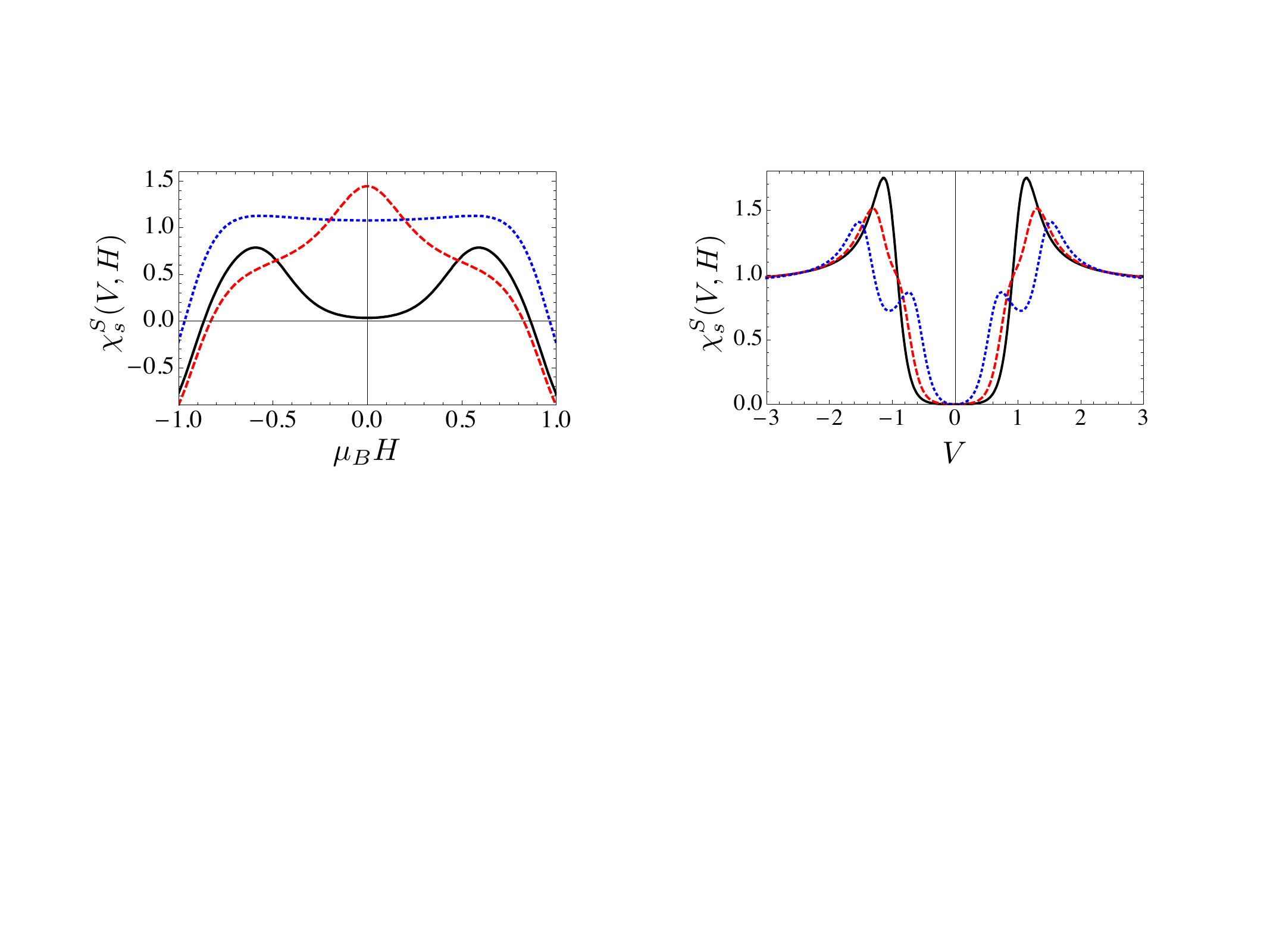}
		\caption{The out-of-equilibrium spin susceptibility $\chi_s^S$. Left: $\chi_s^S$ as a function of the magnetic field $H$ for various values of the applied voltage $V$. The black (full), red (dashed), and blue (dotted) curves correspond to $V=0.5$, $1$, and $2$, respectively. Right: $\chi_s^S$ as a function of the applied voltage $V$ for various values of the applied external field  $H$. The black (full), red (dashed), and blue (dotted) curves correspond to $E_Z=0$, $0.2$, and $0.4$, respectively. Both $E_Z$ and $V$ are expressed in terms of the gap energy $\Delta$. }
	\label{SpinSusc}
\end{figure}

The spin susceptibility of the superconductor is given by:
\begin{equation}
\chi_s^S(V,H)=\frac{\partial S}{\partial H}\,,
\end{equation}
where the spin accumulation $S$ was defined in Eq.~\eqref{accu_eq}. In order to simplify the discussion, we will neglect here any spin accumulation in the leads ($\mu_s^\alpha=0$). We note in passing that for a superconductor at equilibrium, the spin susceptibility is zero for Zeeman splittings $E_Z<\Delta/2$ (and at $T=0$), as it is impossible to create any spin imbalance because of the superconducting gap. In the right plot in Fig.~\ref{SpinSusc} we show $\chi_s^S$ as a function of the Zeeman splitting $E_Z$ for several values of the applied voltage, while in the left plot  we show the spin susceptibility as a function of the  voltage $V$ for various values of the applied field $H$. As expected, the spin susceptibility vanishes as $V$ vanishes. On the other hand, this is finite at zero magnetic field and finite voltage, as the system is very susceptible to build up spin polarization. The spin susceptibility is also a witness of the best strategy in terms of external parameters (i. e. voltage and magnetic field) to magnetize the superconductor.

\section{Results for an applied AC voltage}

\subsection{Time-dependent behavior}
\subsubsection{Numerical results}
We now apply a time-dependent (AC) sinusoidal voltage of frequency $\omega$ and amplitude $V_{rf}$ on the injector, such that $V(t)=V+V_{rf}\textrm{cos}(\omega t)$.
The expressions for the spin currents do not change [see Eq. (\ref{spin_current_FM})]. The only change
comes from the dependence on time of the spin imbalance.  We can solve numerically the self-consistent integral equations of motion [Eqs.~(\ref{motion_eq})-(\ref{id})] to obtain the time-dependent $V_{DET}(t)$, $\mu_s(t)$ and accumulated spin $S(t)$, as a function of applied $V$ for various values of frequencies and AC amplitudes.

\begin{figure}[ht]
	\centering
		\includegraphics[width=0.47\linewidth]{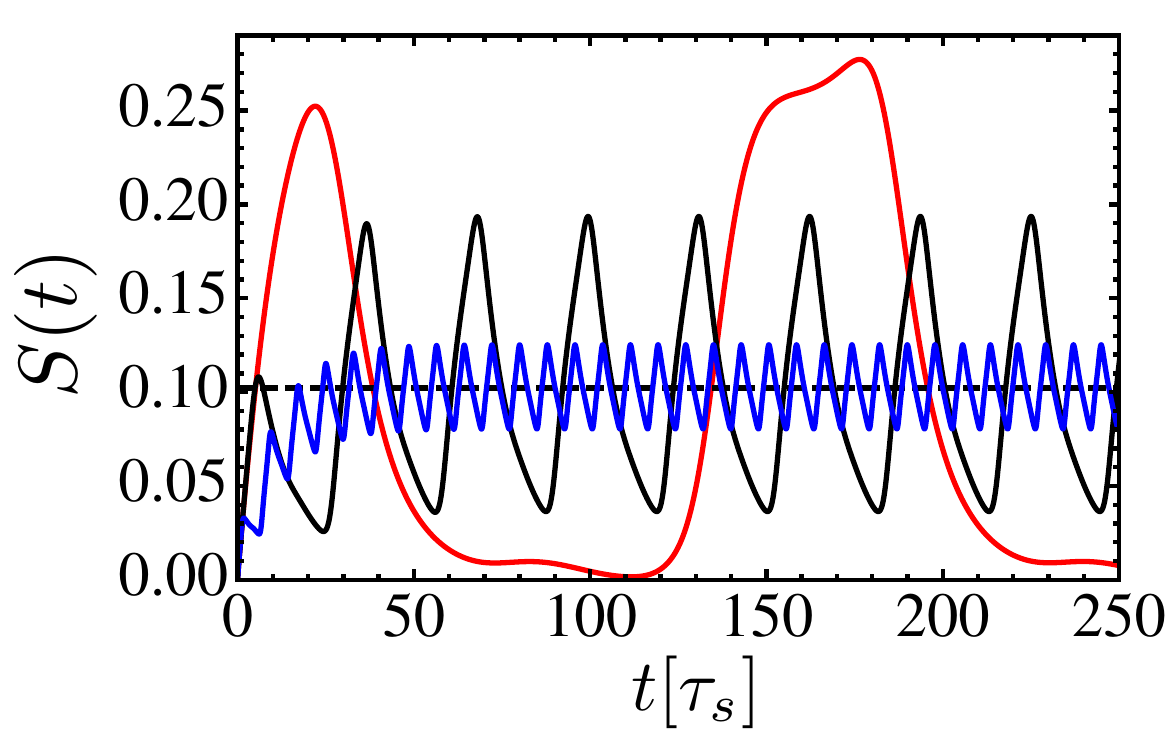}
		\includegraphics[width=0.45\linewidth]{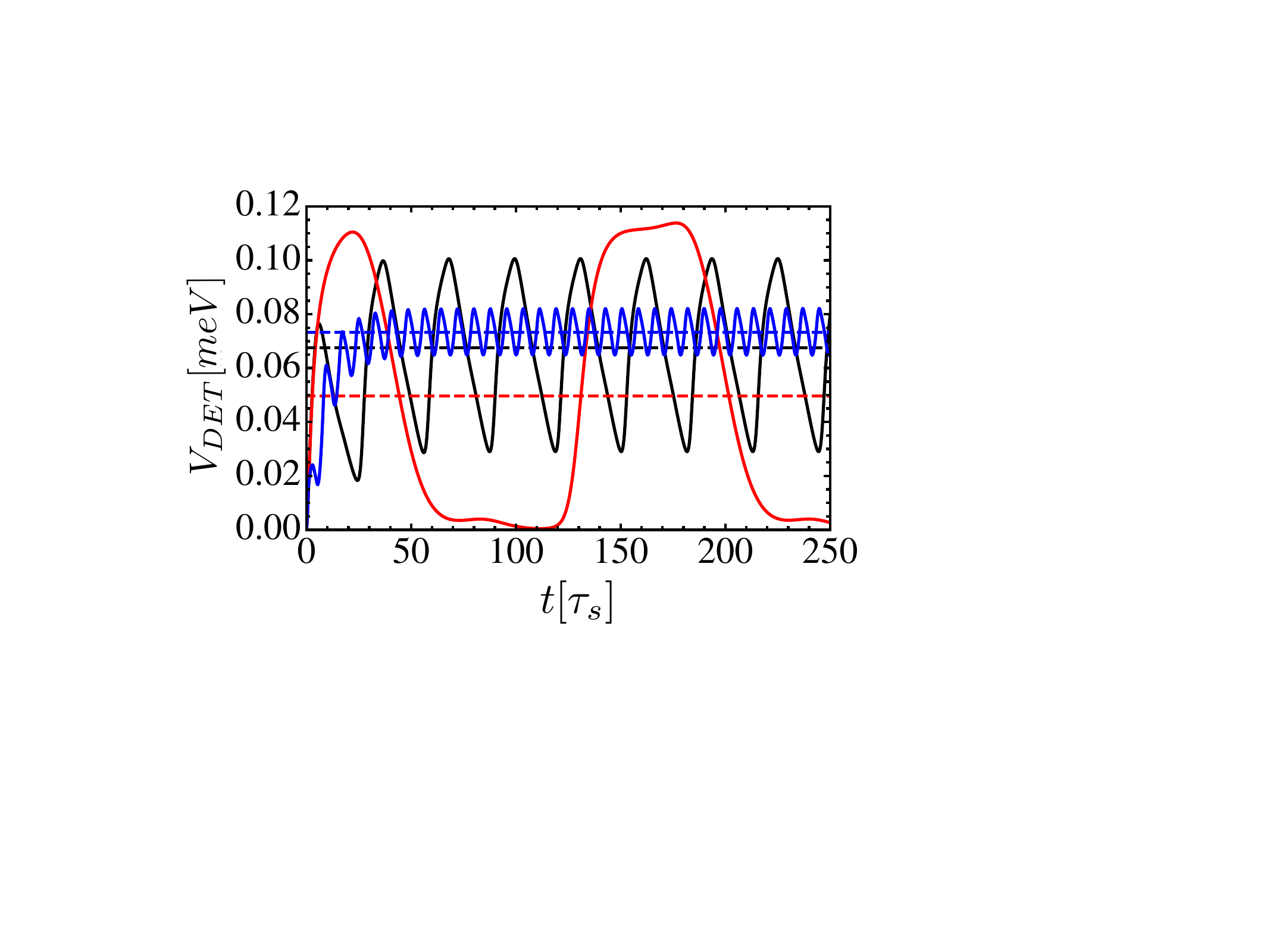}
		\caption{(Color online) Left: Spin imbalance (in arbitrary units) as a function of time (in units of $\tau_s$) Right: Calculated $V_{DET}$ (in $meV$) as a function of time (in units of $\tau_s$). Here, the the parameters are $V_{rf}=0.2meV$, $V=0.1meV$ and $\omega=0.8/\tau_s$ (blue), $\omega=0.2/\tau_s$ (black), and $\omega=0.04/\tau_s$ (red). In the left plot, the average of the oscillations (denoted  by the dotted line) is independent of frequency, while their amplitude is not, while on the right  both the averages (denoted by the corresponding dashed lines) and the amplitude of the oscillations depend on frequency.}
	\label{fig:Fig7a}
\end{figure}

We begin by plotting the accumulated spin $S(t)$ as a function of time.  The time-dependence of the spin accumulation can be understood easily by thinking of the superconductor as a capacitor (its charge could be viewed as the spin imbalance).
Indeed, for large frequencies the capacitor is loading (the spins are accumulating up to a maximal value) but its decreasing never happens because the
spin relaxation time is larger than the period of the oscillations. On the contrary, for small frequencies
voltage the spins can relax because of the large period of the oscillations.
In the left plot in Fig.~\ref{fig:Fig7a} we plot $S(t)$ for three values of frequency, $\omega=0.8/\tau_s$, $\omega=0.2/\tau_s$  and $\omega=0.04/\tau_s$. All other parameters are the same as in the previous section. We note that
the average of the oscillations is independent of frequency, while their amplitude is not. The larger the frequency, the more the behavior of $S(t)$ approaches that of a charging capacitor with smaller and smaller oscillations around the saturation value.

It would thus seem that experimentally one cannot see a frequency dependence for the time-averaged spin accumulation. However, in an actual experiment one does not measure $S$ but $V_{DET}$, which can exhibit a strong non-linear behavior with $S$. We should then expect that if the amplitude of the oscillations in $S(t)$ depends on frequency the time average of $V_{DET}$ depends on frequency via rectification effects. In the right plot in Fig.~\ref{fig:Fig7a} we plot the time dependence of $V_{DET}$ for three different frequencies and we see that indeed both the amplitude of the oscillations and the time average depend on frequency.

\subsubsection{Taylor expansion}

To understand the above numerical results we study a few limiting cases that can be solved analytically.
For a small applied AC voltage ($V_{rf} \ll V$), with $V(t)=V+V_{rf}\textrm{cos}(\omega t)$, we can use a Taylor expansion, and  the spin current can be expressed as
\begin{align}\label{current_expansion}
\I_s(t)&=\I_s[V+V_{rf}\textrm{cos}(\omega t)]\notag\\
&\approx \I_s[V]+\left.\frac{\partial \I_s}{\partial V}\right|_{V}.V_{rf}\textrm{cos}(\omega t)\notag\\
&+\frac{1}{2}\left.\frac{\partial^2 \I_s}{\partial V^2}\right|_{V}.[V_{rf}\textrm{cos}(\omega t)]^2+...
\end{align}
Inserting Eq.~(\ref{current_expansion}) into Eq.~(\ref{imbalance_eq}) gives us an expansion for the spin imbalance in powers of $V_{rf}$ : $S(t)=S_0 (t)+S_1 (t)+S_2 (t)+...$. We focus on the first two terms in the expansion, but the next orders can be studied in a similar fashion. For times much more larger than $\tau_s$ we obtain (see Appendix A).
\begin{align}
&S(t)=\I_s (V)\tau_s\notag\\
&+\left.\frac{\partial \I_s}{\partial V}\right|_{V}.V_{rf}\frac{\tau_s}{1+\tau_s^2\omega^2}\left[\textrm{cos}(\omega t)+\omega\tau_s \textrm{sin}(\omega t)\right].
\end{align}

We see from the above formula that the average accumulated spin is independent of frequency, consistent with the numerical analysis in the previous section (see Fig.~\ref{fig:Fig7a}). However, the amplitude of the oscillations does depend on frequency, with a cutoff/crossover at a frequency $\omega\approx 1/\tau_{s}$. However this analysis in valid only when $V_{rf}$ is very small, and in the regime in which the system is well described by the non-self-consistent calculation. Also, since the timescales involved are very short, it is much harder to have experimentally access to the amplitude of the oscillations than to the time averages, and in what follows we will focus rather on time-averaged quantities than on time-dependent ones. We expect that the non-linearity will give rise to a frequency dependence even when averaging over time, allowing us to detect directly this spin relaxation time in the frequency domain.

\subsection{Time-averaged quantities}

\begin{figure}[ht]
		\includegraphics[width=0.9\linewidth]{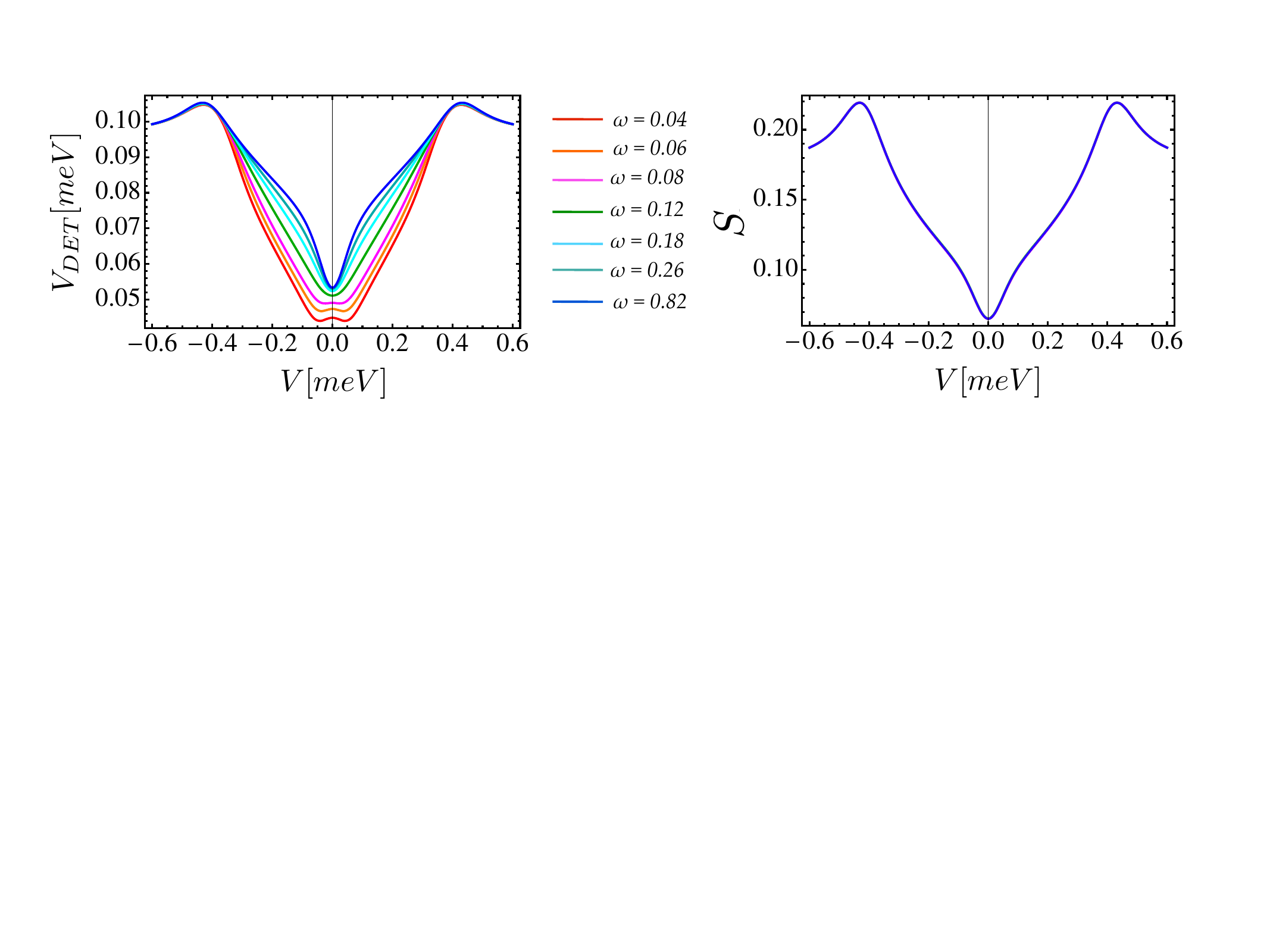}
		\caption{(Color online) Dependence of $V_{DET}$(in $meV$) and $S$ (in arbitrary units) as a function of the applied $V$  (in $meV$) for various frequencies (in units of $1/\tau_s$), at fixed AC amplitude ($V_{rf}=0.2$ meV). Note that all curves overlap on the right plot.}
	\label{fig:Fig4}
\end{figure}

\begin{figure}[ht]
		\includegraphics[width=0.9\linewidth]{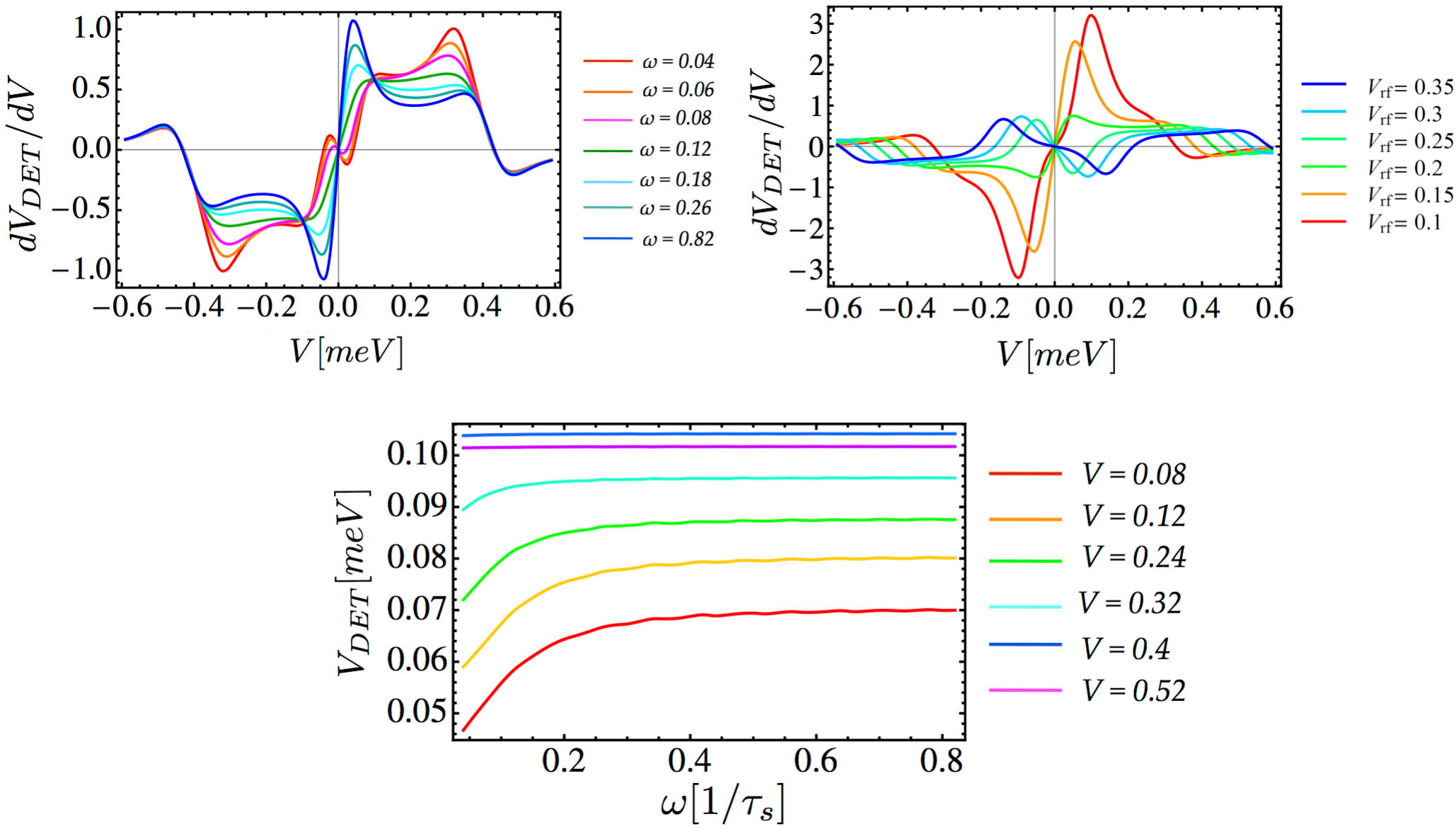}
		\caption{(Color online) Top: Dependence of $d V_{DET}/d V$ (in arbitrary units) as a function of the applied $V$ (in $meV$) for various values of the frequency $\omega$ at fixed $V_{rf}=0.2$ meV) amplitude (left), and for various values of the AC amplitude at a fixed frequency of $\omega=0.2/\tau_s$ (right). Bottom: Dependence of  $V_{DET}$ (in $meV$) on the AC frequency (in units of $1/\tau_s$) at fixed AC voltage amplitude $V_{rf}=0.2meV$, for various applied voltage $V$ (in $meV$).}
	\label{fig:Fig5}
\end{figure}

In general, time-averaged measurements are easier to perform experimentally than time-domain ones. Here we study the dependence of the average measured $V_{DET}$ as a function of the applied voltage for different AC amplitudes and frequencies. We begin by plotting $V_{DET}$ and $S$ as a function of $V$ for different frequencies
 at fixed AC amplitude (see Fig. \ref{fig:Fig4}).  We have checked that while the measured $V_{DET}$ and $\mu_s$ do depend on the frequency (because of the non-linearities in the system), the accumulated spin $S$ does not, as described also in the previous section. We have calculated the accumulated spin $S$ (see right plot in  Fig.~\ref{fig:Fig4}) for different values of $V$ and frequency and we have seen that  $S$ is indeed unaffected by the frequency. All frequencies are given in units of $1/\tau_s$.
Subsequently, in the top plots in Fig.~\ref{fig:Fig5} we show derivative of the average $d V_{DET}/d V$ as a function of applied $V$ for various values of the frequency (left) and  AC amplitudes (right).

We note that the main features that we observe, i.e a flattening of the $V_{DET}$ dependence on $V$, with an eventual extra peak arising at $V=0$, a doubling of the peaks in the $dV_{DET}/dV$ dependence on $V$, whose position depend quasi-linearly on $V_{rf}$, and a saturation of $V_{DET}$ and $dV_{DET}/dV$ with increasing the frequency, are qualitatively similar to what is measured in Ref.~\onlinecite{quay2014}, even if our working assumptions are not necessarily the same. The frequency at which the saturation occurs seems thus to be directly related to the inverse of the spin-relaxation time. 

\subsection{Rectangular voltage pulses}

To get an analytical understanding of the numerical results presented in the previous section we consider also a different type of AC signal, for example a chain of rectangular pulses. In this case we can also calculate analytically the form of the spin imbalance, if we make the assumption that the self-consistent effects are negligible.
The pulse has the following shape
\begin{equation}
V_{rf}(t)=\sum^{N}_{i=0}V_{rf}\Big[\Theta\Big(t-\frac{2\pi i}{\omega}\Big)-\Theta\Big(t-\epsilon-\frac{2\pi i}{\omega}\Big)\Big],
\end{equation}
with $T=2\pi/\omega$ the period of the signal, $V_{rf}$ its amplitude, and $\epsilon$ the width of the pulse. The difference of spin imbalance between the stationary regime ($V_{rf}=0$) and the time-dependent one ($V_{rf}\ne 0$) can be calculated exactly using Eq.~(\ref{imbalance_eq}):
\begin{align}
\delta S(t)&=S[V_{rf}(t)]-S[V_{rf}(t)=0]\notag\\
&=e^{-t/\tau_s}\int^t_0 e^{x/\tau_s} [\I^i_s [V+V_{rf}(x)]-\I^i_S(V)]dx.
\end{align}
Performing the integral over time (see Appendix B) leads to
\begin{eqnarray}\label{accu_time}
\delta S(t)&=&\{\I^i_s (V+V_{rf})-\I^i_S(V)\}\tau_s e^{-t/\tau_s}\Big[(e^{\epsilon/\tau_s}-1)\frac{e^{2\pi N/\omega\tau_s}-1}{e^{2\pi/\omega\tau_s}-1}\nonumber \\
&+&\Theta\Big(t-\frac{2\pi N}{\omega}-\epsilon\Big)\Big(e^{2\pi N/\omega\tau_s+\epsilon/\tau_s}-e^{2\pi N/\omega\tau_s}\Big)
+\Theta\Big(\frac{2\pi N}{\omega}+\epsilon-t\Big)\Big(e^{t/\tau_s}-e^{2\pi N/\omega \tau_s}\Big)\Big].
\end{eqnarray}
The average of Eq. (\ref{accu_time}) can be written as $\delta \bar{S}=1/T \int^T_0 \delta S(t)dt$ and leads to
\begin{equation}
{\delta \bar S}=[\I^i_s (V+V_{rf})-\I^i_S(V)]\frac{\omega\epsilon\tau_s}{2\pi}+{\cal{O}}\Big(e^{-2\pi/\omega \tau_s}\Big).
\label{s1}
\end{equation}

\begin{figure}[ht]
	\centering
		\includegraphics[width=8cm]{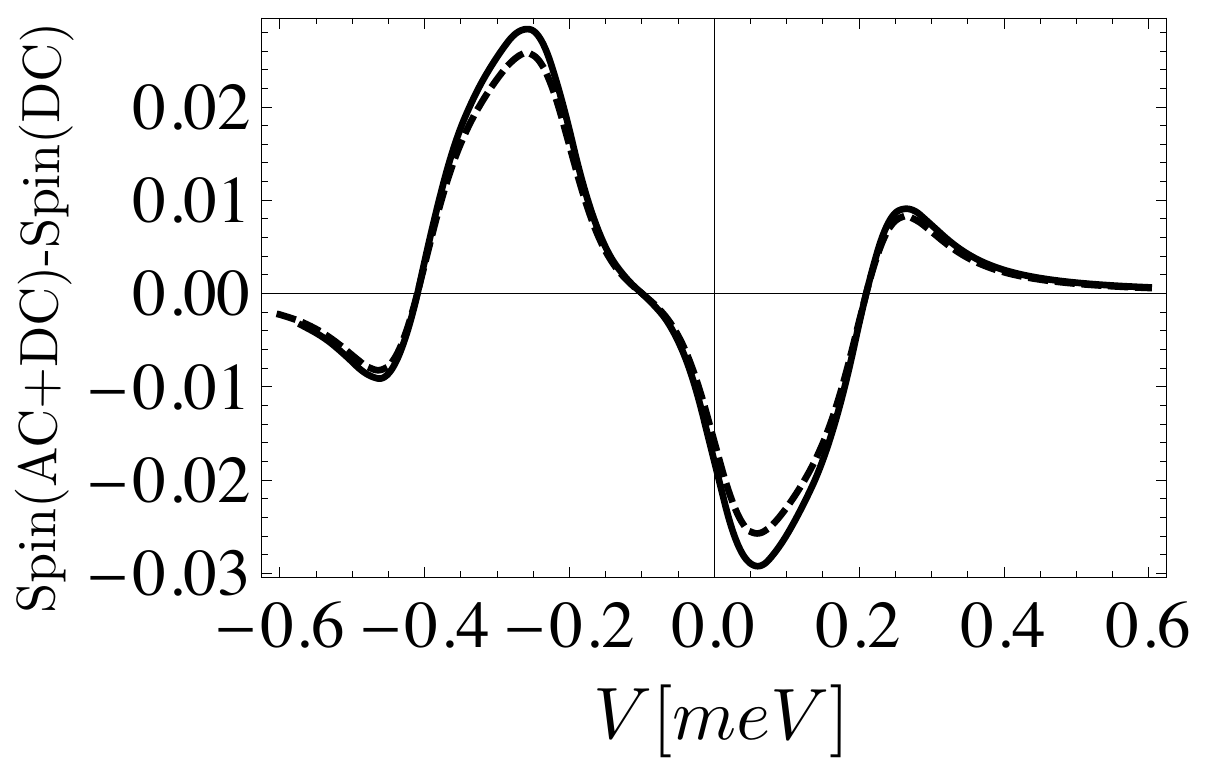}\hspace{0.5cm}
		\includegraphics[width=8cm]{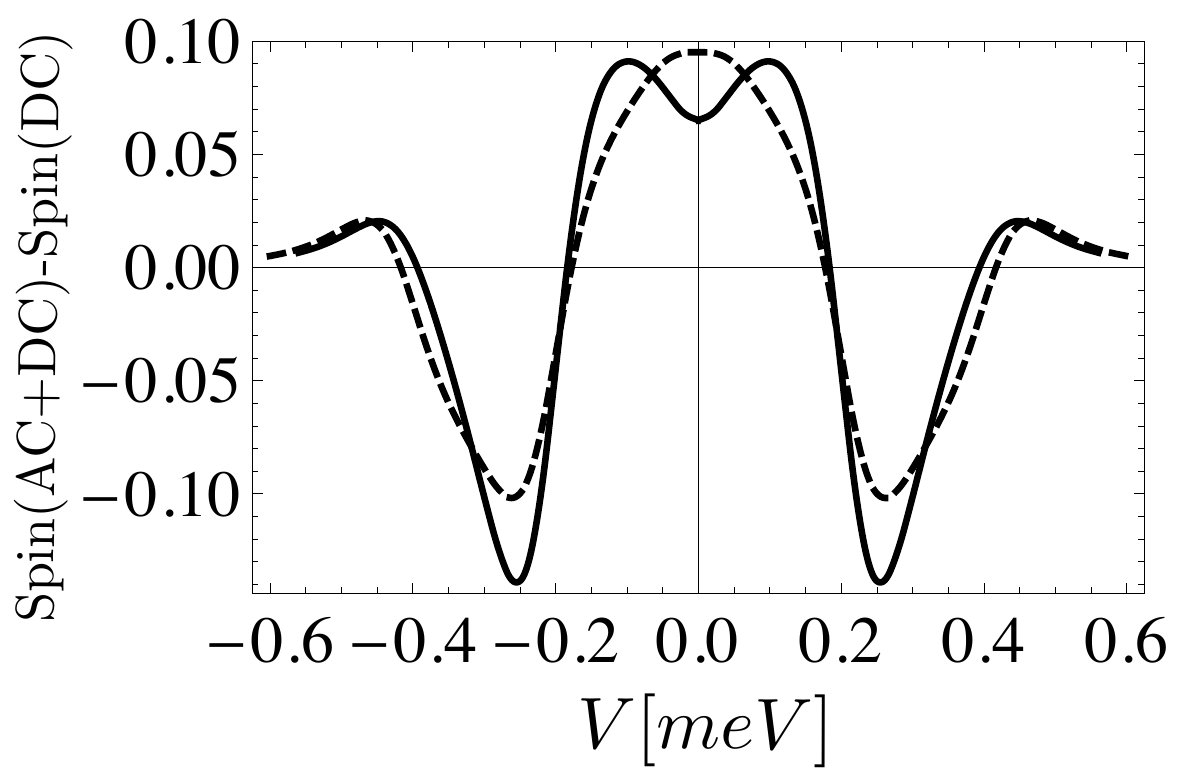}
		\caption{Excess accumulated spin (in arbitrary units) (full line), and $\I^i_s (V+V_{rf})-\I^i_S(V)$(dashed line), as a function of $V$ for $V_{rf}=0.2meV$ and the frequency $\omega=0.04/\tau_s$. Note the qualitatively similar behavior of the two curves (a constant has been introduced to uniformize the two curves).}
	\label{npulse}
\end{figure}

If the pulse constitutes a constant fraction of the period ($\omega \epsilon$ is constant), the accumulated average $S$ should be independent of frequency. We have checked that is indeed the case by a numerical analysis. Also this is consistent with our previous observations for the sinusoidal signal. An interesting observation that we make is that, as it can be seen from Eq.~(\ref{s1}) the dependence of $\delta \bar S$ on $V$ is given generically by $\I^i_s (V+V_{rf})-\I^i_S(V)$. In Fig.~\ref{npulse} we plot the dependence of the excess accumulated spin $\delta \bar S$ as a function of $V$ obtained numerically for a specific value of $V_{rf}$ and of frequency. We also sketch the behavior of $\I^i_s  (V+V_{rf})-\I^i_S(V)$, showing that indeed, to first approximation, the behavior of $\delta \bar S$ follows qualitatively the behavior of  $\I^i_s  (V+V_{rf})-\I^i_S(V)$.

A simple generalization can be made to understand qualitatively the behavior of $\delta \bar S$ with $V$ for the sinusoidal signal. To first approximation a sinusoidal signal is equivalent to a superposition of two $V_{rf}$ and $-V_{rf}$ pulses, with an $\epsilon \omega=1/2$. We would then expect an overall dependence of $\delta \bar S$ qualitatively similar with $\I^i_s  (V+V_{rf})+\I^i_s  (V-V_{rf})-2\I^i_S(V)$.
In Fig.~\ref{npulse} we plot the dependence of the excess accumulated spin as a function of $V$ obtained numerically for a specific value of $V_{rf}=0.2meV$ and $\omega=0.04/\tau_s$ for a sinusoidal signal, and we also sketch the behavior of $\I^i_s  (V+V_{rf})+\I^i_s  (V-V_{rf})-2\I^i_S(V)$, showing that, remarkably enough,  the two behaviors are indeed qualitatively similar.

\subsubsection{Differences between accumulation and relaxation times}

\begin{figure}[ht]
	\includegraphics[width=7cm]{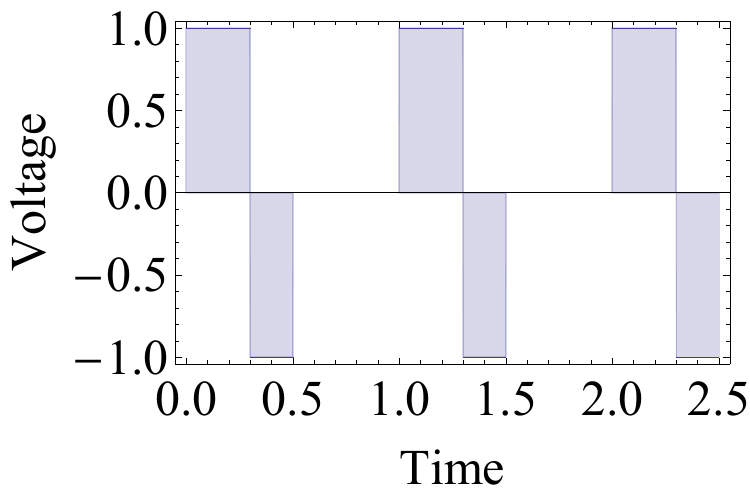}
	\caption{(Color Online) Applied time dependent voltage for $A_1=-A_2=1$, $\epsilon_1=0.3$ and $\epsilon_2=0.2$.}
	\label{fig:Fig8}
\end{figure}

In this section we consider the possibility that the time for accumulation (loading) and relaxation (unloading) are different. Such phenomenon could be detected by applying a time dependent voltage with the following shape (see Fig. \ref{fig:Fig8})
\begin{align}
V_{rf}(t)&=\sum^{N}_{i=0}A_{1}\Big[\Theta\big(t-\frac{2\pi i}{\omega}\Big)-\Theta\Big(t-\epsilon_1-\frac{2\pi i}{\omega}\Big)\Big]\notag\\
&+\sum^{N}_{i=0}A_{2}\Big[\Theta\Big(t-\epsilon_1-\frac{2\pi i}{\omega}\Big)-\Theta\Big(t-\epsilon_1-\epsilon_2-\frac{2\pi i}{\omega}\Big)\Big].
\end{align}
This is because if $A_1>0$ and $A_2<0$ the first step corresponds to the loading of the superconductor and is controlled by $\tau_1$, and the second one to the unloading and is controlled by $\tau_2$. A similar calculation as before can be performed leading to the following form for the average spin imbalance
\begin{align}
\frac{\delta\bar{S}}{\I^i_s (V+V_{rf})-\I^i_S(V)}&=\frac{\omega (A_1\epsilon_1\tau_1+A_2\epsilon_2\tau_2)}{2\pi}.
\end{align}
By setting $\epsilon_1=\epsilon_2=\epsilon$, $A_1=A_2=A$ and $\tau_1=\tau_2=\tau$ we effectively restore  the previous situation
\begin{equation}
\delta \bar{S}=[\I^i_s (V+V_{rf})-\I^i_S(V)]\frac{2\tau\omega\epsilon}{2\pi},
\end{equation}
with $\tau_s$ being replaced by $2\tau=\tau_1+\tau_2$. We can see that in this limit the dependence is still linear with frequency. One important observation to make is that the difference between $\tau_1$ and $\tau_2$ can be measured directly by applying an AC voltage with $A_1=-A_2$ and $\epsilon_1=\epsilon_2$. If the two times are different, the average excess accumulation will be non-zero, which is not the case when $\tau_1=\tau_2$.

\section{Conclusions}

We have calculated the spin accumulation induced in a SC in the presence of a Zeeman field in an NS junction taken out of equilibrium, as well as the possibility to measure it non-locally using a second ferromagnetic probe. We have found that for an applied DC voltage, the dependence of the non-local signal has the same qualitative behavior as the BCS NS non-linear conductance, i.e. a reduced value at low voltages and peaks at voltages close to the SC gap. Most importantly, we have shown that in the presence of an AC voltage the time average of the non-local signal is frequency dependent, and we have noted that this is mainly due to the non-linear response of the detector. Our theoretical predictions for such frequency dependence show a very good qualitative agreement with the experimental measurements in Ref. 15, and may allow one to estimate experimentally the value of the spin-relaxation time. Also we have studied the effect of the spin accumulation in the normal leads and we have calculated the out-of-equilibrium spin susceptibility of the SC, which we have shown to be very different from its equilibrium value.

\acknowledgments We thank J. Gabelli, J. S. Meyer and M. Houzet for helpful discussions. This work was funded by an ERC Starting Independent Researcher Grant NANOGRAPHENE 256965; an ERC Synergy grant; an ANR Blanc grant (MASH) from the French Agence Nationale de Recherche; an ANR JCJC grant (SPINOES) from the French Agence Nationale de Recherche and the Netherlands Organization for Scientific Research (NWO/OCW).

\vspace{1cm}

\noindent
{\it \bf Appendix A - Taylor expansion of $S(t)$.}

\begin{flushleft}
\underline{Calculate $S_0 (t)$}
\end{flushleft}
\begin{align}
S_0 (t)&=\left[\int_0^{t} e^{t'/\tau_s} \I_s (V) dt' \right]e^{-t/\tau_s}\notag\\
&=\tau_s \I_s (V) (e^{t/\tau_s}-1)e^{-t/\tau_s}\notag\\
&=\I_s (V)\tau_s -\tau_s e^{-t/\tau_s} \I_s (V)
\end{align}
where the second term corresponds to the transient  term which vanishes after $t\gg\tau_s$. The average of $S_0 (t)$ over one period yields $S_M^{(0)}=\I_s (V)\tau_s$ which does not depend on $\omega$.
\begin{flushleft}
\underline{Calculate $S_1 (t)$}
\end{flushleft}
\begin{align}\label{first_spin}
S_1 (t)&=\left.\frac{\partial \I_s}{\partial V}\right|_{V}.\left[\int_0^{t} e^{t'/\tau_s} V_{rf}\textrm{cos}(\omega t') dt' \right]e^{-t/\tau_s}\notag\\
&=\left.\frac{\partial \I_s}{\partial V}\right|_{V}.V_{rf}e^{-t/\tau_s}\int_0^{t}e^{t'/\tau_s}\textrm{cos}(\omega t')dt'\notag\\
&=\left.\frac{\partial \I_s}{\partial V}\right|_{V}.V_{rf}e^{-t/\tau_s}\frac{\tau_s}{1+\tau_s^2\omega^2}\left\{-1\right.\notag\\
&+\left.e^{t/\tau_s}\left[\textrm{cos}(\omega t)+\omega\tau_s \textrm{sin}(\omega t)\right]\right\}\notag\\
&=\left.\frac{\partial \I_s}{\partial V}\right|_{V}.V_{rf}e^{-t/\tau_s}\frac{-\tau_s}{1+\tau_s^2\omega^2}\notag\\
&+\left.\frac{\partial \I_s}{\partial V}\right|_{V}.V_{rf}\frac{\tau_s}{1+\tau_s^2\omega^2}\notag\\
&\ \ \ \ \ \ \ \ \ \ \ \ \ \ \ \ \ \ \ \ \ \ \ \ \ \ \ \ \ .\left[\textrm{cos}(\omega t)+\omega\tau_s \textrm{sin}(\omega t)\right]\notag\\
&=\left.\frac{\partial \I_s}{\partial V}\right|_{V}.V_{rf}\frac{\tau_s}{1+\tau_s^2\omega^2}\notag\\
&\ \ \ \ \ \ \ \ \ \ \ \ \ \ \ \ \ \ \ \ \ \ \ \ \ \ \ \ \ .\left[\textrm{cos}(\omega t)+\omega\tau_s \textrm{sin}(\omega t)\right].\notag\\
\end{align}
Note the dependence on frequency of the prefactor $\frac{\tau_s}{1+\tau_s^2\omega^2}$, corresponding to a frequency dependence for the amplitude of the oscillations in $S(t)$.

\vspace{1cm}

\noindent
{\it \bf Appendix B - Analytical calculation of $S(t)$ for a rectangular pulse.} \\

Eq. (\ref{imbalance_eq}) can be integrated analytically if the applied AC signal is a rectangular pulse. This yields
\begin{align}
\delta S(t)&=[\I^i_s (V+V_{rf})-\I^i_S(V)]e^{-t/\tau_s}\left\{\sum^{N-1}_{i=0}\int^{2\pi i/\omega+\epsilon}_{2\pi i/\omega}e^{x/\tau_s}dx\right.\notag\\
&+\int^{2\pi N/\omega+\epsilon}_{2\pi N/\omega}e^{x/\tau_s}\Theta\left(t-\frac{2\pi N}{\omega}-\epsilon\right)\notag\\
&+\int^{t}_{2\pi N/\omega}e^{x/\tau_s}\Theta\left(\frac{2\pi N}{\omega}+\epsilon-t\right).
\end{align}

Performing those integrals leads to
\begin{align}
\delta S(t)&=[\I^i_s (V+V_{rf})-\I^i_S(V)]\tau_s e^{-t/\tau_s}\left\{\left(e^{\epsilon/\tau_s}-1\right)\frac{e^{2\pi N/\omega \tau_s}-1}{e^{2\pi/\omega \tau_s}-1}\right.\notag\\
&+\Theta\left(t-\frac{2\pi N}{\omega}-\epsilon\right)\left(e^{2\pi N/\omega \tau_s+\epsilon/\tau_s}-e^{2\pi N/\omega \tau_s}\right)\notag\\
&+\left.\Theta\left(\frac{2\pi N}{\omega}+\epsilon-t\right)\left(e^{t/\tau_s}-e^{2\pi N/\omega \tau_s}\right)\right\}.
\end{align}
The last part of the calculation is the average over one period of the oscillations, $\delta \bar{S}=1/T\int^T_0 \delta S(t) dt$, which leads to
\begin{align}
\delta \bar{S}&=[\I^i_s (V+V_{rf})-\I^i_S(V)]\frac{\omega\tau_s}{2\pi}\int^{2\pi(N+1)/\omega}_{2\pi N/\omega}dt\notag\\
&\ \ \ \ \ \ \ \ \ \ \ \ \ \ \ \ \ \ \ \ \ \ \ \ \ \ \ \ \ \ \ \ \ .\left(e^{\epsilon/\tau_s}-1\right)\frac{e^{2\pi N/\omega \tau_s}-1}{e^{2\pi/\omega \tau_s}-1}e^{-t/\tau_s}\notag\\
+&[\I^i_s (V+V_{rf})-\I^i_S(V)]\frac{\omega\tau_s}{2\pi}\int^{2\pi N/\omega+\epsilon}_{2\pi N/\omega}dte^{-t/\tau_s}\left(e^{t/\tau_s}-e^{2\pi N/\omega \tau_s}\right)\notag\\
+&[\I^i_s (V+V_{rf})-\I^i_S(V)]\frac{\omega\tau_s}{2\pi}\int^{2\pi (N+1)/\omega}_{2\pi N/\omega+\epsilon}dt\notag\\
&\ \ \ \ \ \ \ \ \ \ \ \ \ \ \ \ \ \ \ \ \ \ \ \ \ \ \ \ .e^{-t/\tau_s}\left(e^{2\pi N/\omega \tau_s+\epsilon/\tau_s}-e^{2\pi N/\omega \tau_s}\right),
\end{align}
which gives
\begin{align}
\delta \bar{S}&=[\I^i_s (V+V_{rf})-\I^i_S(V)]\tau_s\left\{\frac{\omega\tau_s}{2\pi}e^{-2\pi/\omega\tau_s}\left(e^{\epsilon/\tau_s}-1\right)\right.\notag\\
+&\frac{\omega\epsilon}{2\pi}+\frac{\omega\tau_s}{2\pi}\left(e^{-\epsilon/\tau_s}-1\right)\notag\\
+&\left.\frac{-\omega\tau_s}{2\pi}\left(e^{-2\pi/\omega\tau_s}-e^{-\epsilon/\tau_s}\right)\left(e^{\epsilon/\tau_s}-1\right)\right\}.
\end{align}
Summing all contributions give us the final result for the average spin imbalance
\begin{equation}
\delta \bar{S}=[\I^i_s (V+V_{rf})-\I^i_S(V)]\frac{\tau_s\omega\epsilon}{2\pi}+{\cal{O}}(e^{-2\pi N/\omega\tau_s}).
\end{equation}

\end{document}